\documentclass[%
 reprint,
 amsmath,amssymb,
 aps,
]{revtex4-1}

\usepackage{graphicx}%
\usepackage{dcolumn}%
\usepackage{bm}%
\usepackage{hyperref}%

\usepackage{siunitx}

\usepackage{color}
\definecolor{commentColorWTJ}{rgb}{0.6,0.2,0.0}
\definecolor{commentColorASN}{rgb}{0.2,0.6,0.0}
\definecolor{commentColorRNP}{rgb}{0.0,0.0,1.0}
\definecolor{commentColorRVL}{rgb}{0.0,0.5,0.5}

\begin{document}

\preprint{APS/123-QED}

\title{Nanobenders: efficient piezoelectric actuators for widely tunable nanophotonics at CMOS-level voltages}%

\author{Wentao Jiang}
\thanks{These authors contributed equally to this work}
\author{Felix M. Mayor}
\thanks{These authors contributed equally to this work}
\author{Rishi N. Patel}
\author{Timothy P. McKenna}
\author{Christopher J. Sarabalis}
\author{Amir H. Safavi-Naeini}%
\email{safavi@stanford.edu}

\affiliation{Department of Applied Physics and Ginzton Laboratory, Stanford University, 348 Via Pueblo Mall, Stanford, California 94305, USA}

\date{\today}%

\begin{abstract}
Tuning and reconfiguring nanophotonic components is needed to realize systems incorporating many components. The electrostatic force can deform a structure and tune its optical response. Despite the success of electrostatic actuators, they suffer from trade-offs between tuning voltage, tuning range, and on-chip area. Piezoelectric actuation could resolve all these challenges. Standard materials possess piezoelectric coefficients on the order of $\sim\SI{0.01}{\nano\meter/\volt}$, suggesting extremely small on-chip actuation using potentials on the order of one volt. Here we propose and demonstrate compact piezoelectric actuators, called nanobenders, that transduce tens of nanometers per volt. 
By leveraging the non-uniform electric field from submicron electrodes, we generate bending of a piezoelectric nanobeam. Combined with a sliced photonic crystal cavity to sense displacement, we show tuning of an optical resonance by $\sim\SI{5}{\nano\meter/\volt}~(\SI{0.6}{\tera\hertz/\volt})$ and between $1520\sim 1560~\SI{}{\nano\meter}$ ($\sim 400$ linewidths) with only $ \SI{4}{\volt} $. Finally, we consider other tunable nanophotonic components enabled by  nanobenders.
\end{abstract}

\maketitle

Complete and low-power control over the phase and amplitude of light fields remains a major challenge in integrated photonics. Optical components providing such control are essential in systems being developed for optical computing, signal processing, sensing, and imaging~\cite{miller2017attojoule,hamerly2019large,pai2019parallel}. Tuning the optical response of an element entails changing its refractive index by, for example, modifying its temperature, imposing electric fields, or mechanically deforming it. Among these, mechanical deformations have the advantage of being essentially lossless, requiring no static power consumption, and possessing an enormous tuning range and cryogenic compatibility~\cite{midolo2018nano,Safavi-Naeini2019}. Nano-opto-electro-mechanical (NOEM) devices~\cite{zheludev2016reconfigurable,midolo2018nano} have thus been pursued and demonstrated, to realize switches and couplers for classical and quantum light~\cite{van2012ultracompact,han2015large, seok2016large, papon2019nanomechanical,Haffner2019}, resonant and static electro-optomechanical tuning and electro-optical transduction~\cite{perahia2010electrostatically,winger2011chip,bagci2014optical,andrews2014bidirectional,pitanti2015strong,grutter2018invited,bekker2018free}.

An efficient NOEM device solves two problems simultaneously. It is an optical device whose properties are exceptionally sensitive to mechanical deformations. It is also an electromechanical device where a modest voltage can induce large deformations. State-of-the-art optomechanical cavities routinely achieve coupling coefficients $g_\text{OM}/2\pi$ in excess of $\SI{100}{\giga\hertz/\nano\meter}$, and largely satisfy the former requirement. It is more-so the latter requirement of large voltage-induced displacement, which has remained a formidable challenge in this context. Here, two approaches present themselves: electrostatic and piezoelectric forces. Electrostatic forces are generated by the voltage-induced polarization in a material. They do not require any special material property and have been previously used to implement a variety of NOEM systems. However, electrostatic tuning is limited in terms of the achievable tunability and sensitivity due to a trade-off between the generated forces (inversely proportional to capacitor plate spacing) and tuning range (proportional to capacitor plate spacing). Moreover, the quadratic relationship between the induced force and the voltage, and pull-in effect complicate tuning. The piezoelectric effect, which relies on the built-in polarization of a material, has the potential to address all these challenges. There, the displacement-voltage relationship is linear and bidirectional, and is free from the pull-in effect and force/range trade-off. 
This has led to efforts~\cite{hosseini2015stress, tian2018unreleased, jin2018piezoelectrically, stanfield2019cmos} to implement piezo-optomechanically tunable cavities and waveguides and resulted, for example, in demonstrations of cavity wavelength tuning coefficients ranging from 0.1 to 30 pm/V~\cite{tian2018unreleased, jin2018piezoelectrically, stanfield2019cmos}.

In this work, by considering the interplay between non-uniform electric fields and transverse components of the piezoelectric tensor $\bm d$, we discover an actuation mechanism specific to nanoscale piezoelectric actuators that leads to a two-order-of-magnitude increase in achievable displacements. 
We propose and demonstrate a compact ($\sim \SI{10}{\micro\meter}^2$) and geometrically isolated actuator, which we call a ``nanobender'', composed of monolithic metal electrodes on a single layer of a thin-film piezoelectric. The displacement of the nanobender scales quadratically with its length $L$ and can be as large as $\SI{20}{\nano\meter/\volt} $ for $L \sim \SI{20}{\micro\meter}$. %

The enormous sensitivity and tuning range achieved in these nanobenders allow us to achieve a significant breakthrough in NOEM performance. We demonstrate a ``zipper'' optomechanical cavity~\cite{eichenfield2009picogram,leijssen2015strong} actuated by four nanobenders that deform the structure to tune the optical resonance wavelength by $\sim \SI{5}{\nano\meter/\volt}$. With a tuning speed approaching $ \SI{1}{\mega\hertz}$, and a tuning range of $ \SI{60}{\nano\meter}$ with around $\SI{4}{\volt}$, we show single-mode tuning across the full telecom C-band with a CMOS voltage. We further show that the displacement generated by the nanobenders is sufficiently large to ``zip'' and ``unzip'' the zipper cavity, reversibly manipulating the \emph{mechanical} mode structure of nanomechanical resonators with switchable contact forces.

\section*{Results}

\begin{figure*}[tb]
  \includegraphics[scale=0.9]{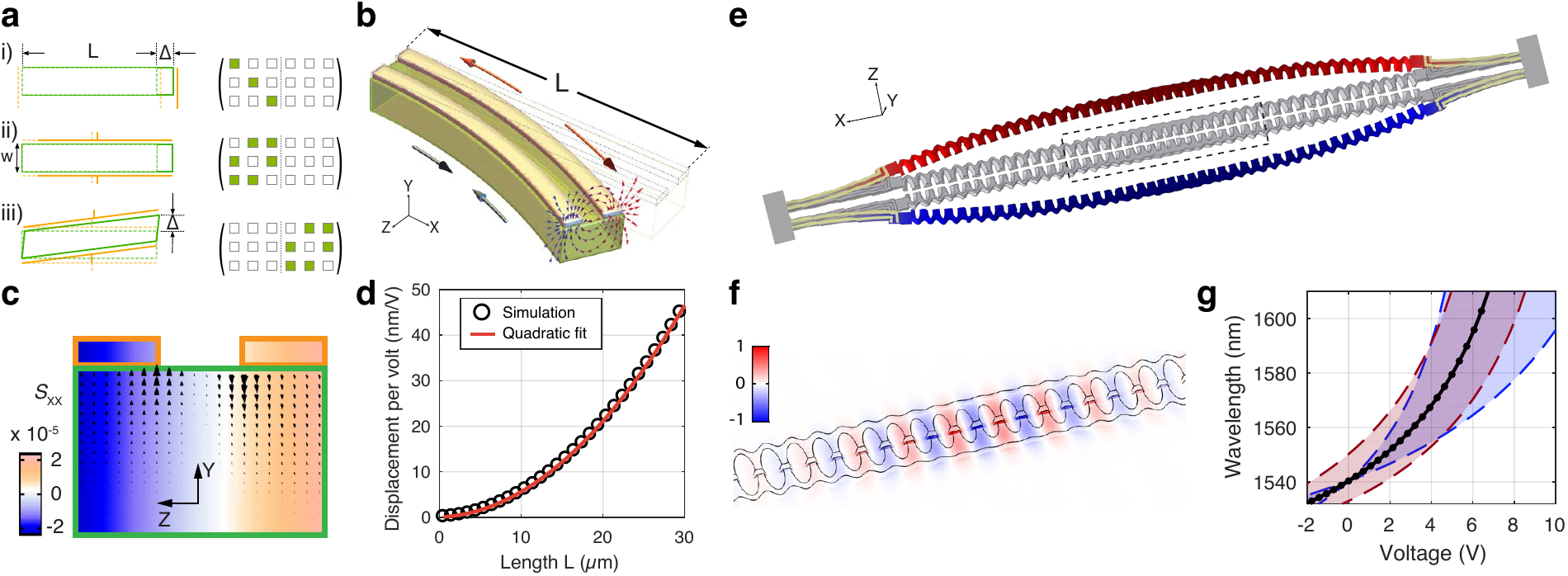} %
\caption{\label{Fig1:nanobender}\textbf{The nanobender and tunable bender-zipper cavity design. a}, Schematic representation of how the components of the piezoelectric tensor affect the deformation of a beam (green) when a voltage is applied on electrodes (orange).  \textbf{b}, Visualization of a nanobender showing an applied electric field (small arrows), resulting strain (large arrows) and displacement of the fixed beam.  \textbf{c}, Cross-section of a nanobender showing the strain $S_{XX}$ as well as the electric field $E_Y$ (arrow-heads) generated by the metal electrodes. \textbf{d}, Simulated displacement at the tip of a nanobender as a function of its length. Notice the quadratic relationship. \textbf{e}, Displacement $u_Z$ of the bender-zipper cavity with a voltage applied to four $L=\SI{5}{\micro\meter}$ nanobenders. The deformation is amplified for illustration purposes. \textbf{f}, Normalized $E_Z$ component of the fundamental optical mode of the zipper cavity. \textbf{g}, Optical resonance wavelength vs voltage applied on a $L=\SI{15}{\micro\meter}$ bender (black curve) for an initial gap of $203$ nm. The blue region covers a bender length variation of $\pm \SI{5}{\micro\meter}$. The red region covers initial gap sizes of $\pm 50$ nm.}
\end{figure*}

\textbf{Operating principle of a piezoelectric nanobender.} Consider a slab of piezoelectric material sandwiched between two electrodes that are separated by a length $L$ and have  a potential difference of $U$ (Fig.~\ref{Fig1:nanobender}a-i). The piezoelectric property of a material is represented by its charge piezoelectricity tensor $\bm d$, a third-rank tensor that relates strain to electric field inside a material ($\bm S = \bm d \cdot \bm {E}$). The terms $d_{11(22,33)}$ (Voigt notation) couple $S_{kk}$ to $E_k$, for $k=x,y,z$, causing compressional/tensile strain to build up in the direction of the electric field. This leads to a displacement $\Delta = S L = d_{11} U$. Considering that $d\approx10~\text{pm/V}$ in standard piezoelectric materials such as aluminum nitride (AlN) and lithium niobate (LN)~\cite{weis1985lithium,guy1999extensional}, such a transducer would only generate displacements at the atomic scale for voltages $U\sim\SI{1}{\volt}$ which are easily produced by CMOS circuits. 

The above expression $\Delta=d_{11} U$ also shows that the generated displacement does not depend on the size of the transducer. This applies to any piezoelectric actuator. We write the constitutive relation between the strain and the electric field $\bm S = \bm d \cdot \bm {E}  $, as $ ((\bm \nabla \bm u)^T +\bm \nabla \bm u)/2 = \bm d \cdot \bm \nabla U $ where $ \bm u(\bm r)$ is the displacement field and $ U(\bm r) $ is the electric potential distribution for a given actuator geometry, applied voltage and boundary conditions. If the geometry of both the actuator and the boundary conditions are scaled by a factor $\epsilon$ while keeping the same applied voltage, then $\bm u'(\bm r) = \bm u( \bm r/\epsilon)$ and $U'(\bm r) = U( \bm r/\epsilon) $ are solutions to the new equations. In other words, the magnitude of the displacement stays constant as the actuator is shrunk leading to an increase in  relative displacement that favors smaller actuators.

As illustrated above, the diagonal elements of $\bm d$, $d_{11(22,33)}$ give rise to tens of picometers of displacement at a potential of around one volt. A much larger displacement can be generated with transverse $\bm d$ ($d_{12(13,23)}$) components (Fig.~\ref{Fig1:nanobender}a-ii). In this situation, the potential $U$ gives rise to an electric field $U/w$ across the width $w$, which generates strain along the length $L$ of the beam. This  leads to a displacement $\Delta \approx d\cdot U L/w \sim (\SI{0.01}{\nano\meter/\volt})\cdot \mathcal R U $, where we have defined the aspect ratio of the actuator $\mathcal R \equiv L/w$. Compared to the previous case, the displacement is enhanced by $\mathcal R $. However, reaching $\Delta \sim \SI{100}{\nano\meter}$ with one volt still requires $ \mathcal R =10^4$, roughly on the same order as that of a long strand of human hair, or sheet of paper, making it impractical.

Is there a configuration that results in a displacement which scales faster than linear with $\mathcal R $? Bending of a beam generates a displacement proportional to $L^2$, where contraction occurs in one half of the beam and expansion in the other half. Looking back to the corresponding electric field $E$, we recognize that bending can be actuated by flipping the direction of $E$ across the width of the beam. Assuming that the derivative of the $E$ field is constant across the width of the beam $\partial_z E \sim U/w^2$ ($z$ is transverse to the beam), the end-point displacement can be approximated by (supplementary information)
\begin{equation}
    \label{eq:disp-length}
	\Delta  \approx \frac{1}{2} d \cdot \partial_{z} E \cdot L^2 \sim (\SI{0.01}{\nano\meter/\volt})\cdot \mathcal{R}^2 U.
\end{equation}
 We see that the displacement in this case is enhanced by the square of the aspect ratio $\mathcal{R}^2$. The required $\mathcal R$ for $\Delta \sim \SI{100}{\nano\meter}$ is drastically decreased to a practical value $ \mathcal R = 100$. As an example, such a non-uniform $E$ field on a $w\sim \SI{400}{\nano\meter} $ wide beam with length $L\sim \SI{40}{\micro\meter} $ would enable actuation of $ \SI{100}{\nano\meter} $ displacement per volt -- a displacement on the same order as the width of the beam. We emphasize that the nanoscale aspect of the nanobender is important for achieving such a large relative displacement. By the scale-invariance arguments above, a larger structure would generate the same displacement, leading to a less appreciable relative motion.  %

A strongly inhomogeneous $E$ field is naturally generated by the fringing fields of a submicron-scale electrode configuration. We consider a simple device, which we call the nanobender, where a pair of parallel electrodes lies on the top surface of a beam made of a thin piezoelectric LN film.  For a beam oriented along crystal axis $X$, the inhomogeneous $E_Y$ field induces a varying strain $S_{XX}$ and results in bending of the beam (Fig.~\ref{Fig1:nanobender}b) via the piezoelectric tensor element $d_{21}$. For $Y$-cut LN where the crystal $Y$ axis is perpendicular to the chip, this bending gives rise to an in-plane displacement that scales quadratically with $L$. Finite-element simulations~\cite{COMSOL} of the nanobender are shown in Fig.~\ref{Fig1:nanobender}c. The simulated $E_Y$ (arrowheads) changes sign along $Z$, causing expansion in one half of the beam and contraction in the other. For all the simulations and experiments, we use a nanobender width $w = 450$ nm, LN thickness $t = 300$ nm, electrode width $w_\text{m} = 150$ nm, electrode-electrode gap $w_\text{g} = 150$ nm  and an aluminum electrode thickness $t_\text{m} = 50$ nm. A more detailed study of how these parameters affect nanobender performance is presented in the supplementary information. Once the nanobender's cross section is fixed, the length $L$ ultimately determines the maximum displacement $\Delta $ generated at the end of the beam. Through simulations (Fig.~\ref{Fig1:nanobender}d) we are able to confirm the quadratic length-displacement relationship in equation~(\ref{eq:disp-length}). The simulated displacement along the other two directions are more than one order of magnitude smaller.

Actuation that induces bending is commonly adopted by macroscopic piezoelectric actuators, realizing displacement per volt values similar to the nanobenders ($\sim\SI{10}{\micro\meter} $ with $\sim \SI{100}{\volt}$)~\cite{safari2008piezoelectric}. In these actuators, a non-uniform strain distribution is achieved by combining multiple layers of different materials, some of which are piezoelectric. However, such an approach is impractical at the nanoscale and difficult to realize in an integrated platform, especially for in-plane actuation. Remarkably,  electrostatic forces can also be used to generate bending with large travel~\cite{conrad2015small}, though scaling actuators down to a few microns is challenging, and current demonstrations require much larger footprints for similar displacements ($\sim 50\times 2000~\SI{}{\micro\meter}^2$ for $\sim \SI{0.1}{\nano\meter/\volt}^2 $).

\begin{figure*}[t]
  \includegraphics[scale=0.95]{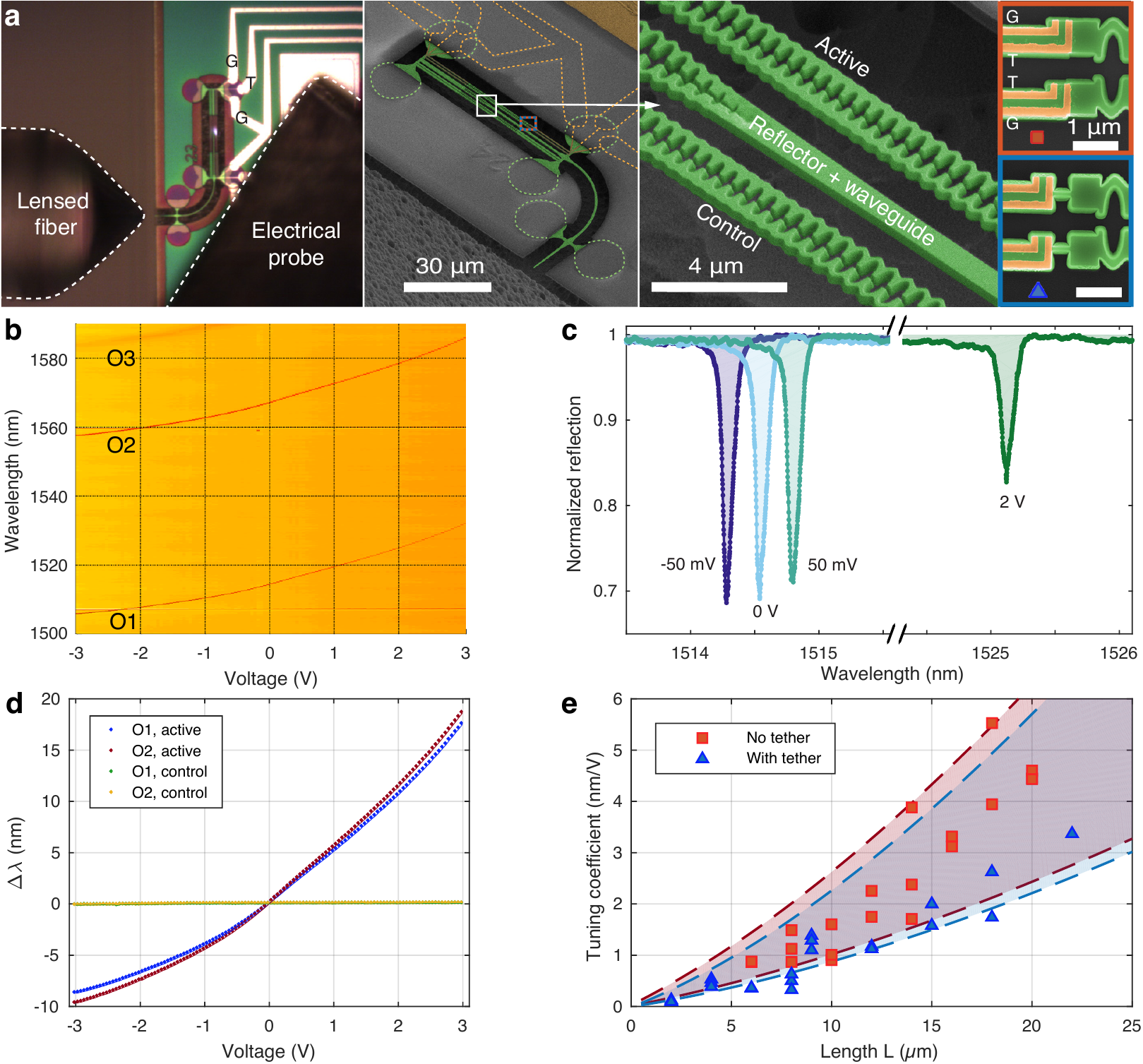} 
\caption{\label{fig:DCtuning}\textbf{DC tuning of a bender-zipper cavity. a}, Left: optical microscope image showing second harmonic generation happening in the zipper cavity as well as the coupling fiber and electrical probe tip. Middle and right: false color scanning electron micrographs of the entire device, the waveguide-zipper coupling region and nanobenders attached to the zipper cavity with (blue triangle) and without tether (red square). \textbf{b}, Measured DC tuning of three optical modes of a single zipper cavity attached to four $L=\SI{15}{\micro\meter}$ nanobenders through narrow tethers. The background is removed through normalization (supplementary information). \textbf{c}, Selected cut-lines from Fig.~\ref{fig:DCtuning}b at $-50$ mV, $0$ V, $50$ mV and $2$ V for the fundamental optical mode O1. \textbf{d}, Wavelength tuning as a function of voltage for two modes of the active and control devices. The data is extracted from Fig.~\ref{fig:DCtuning}b. The control device does not tune since it does not possess nanobenders. \textbf{e}, Measured tuning coefficient versus nanobender length $L$. The dashed lines correspond to simulations with initial gaps of $108$ and $258$ nm.}
\end{figure*}

\textbf{Integration with a zipper cavity.} By integrating the nanobender with a nanophotonic ``zipper'' cavity  \cite{eichenfield2009picogram,leijssen2015strong} on the thin-film LN material platform, we demonstrate its potential for realizing photonic devices with wide low-voltage tunability. A zipper cavity is a sliced photonic crystal consisting of two nano-patterned beams separated by a gap $\tilde{g}_0 \sim \SI{200}{\nano\meter}$ that confines an optical resonance. The $E_Z$ component of the fundamental optical cavity mode is plotted in Fig.~\ref{Fig1:nanobender}f. Due to the sub-wavelength confinement of the mode, the resonance wavelength of the cavity is strongly dependent on the gap between the two beams (supplementary information). A voltage applied to the nanobenders moves the two halves of the zipper cavity (Fig.~\ref{Fig1:nanobender}e), tuning the optical resonance wavelength. In Fig.~\ref{Fig1:nanobender}g we present the simulated voltage-wavelength tuning curve. The tuning curve is nonlinear due to the large changes in $\tilde{g}_0$ -- a smaller $\tilde{g}_0$ increases the optical mode confinement and optomechanical coupling, increasing the slope of the tuning curve.\\

To couple light into the device, we use an edge coupling scheme where a lensed fiber is aligned to a tapered waveguide (Fig.~\ref{fig:DCtuning}a). Light is guided to a reflector and evanescently couples to both an active and a control zipper cavity. The reflection spectrum of the device is recorded for all subsequent measurements (see methods). The bender-zipper cavity is positioned such that the nanobenders are parallel to the crystal $X$ axis, necessary for the nanobenders to operate as designed. We also fabricate and measure a device with nanobenders perpendicular to the crystal $X$ axis and measure two orders of magnitude lower tuning (see supplementary information). We attach the nanobenders to the zipper cavity with and without narrow tethers and measure larger tuning in the untethered devices (highlighted in blue and red in Fig.~\ref{fig:DCtuning}a,e). To apply a voltage to the nanobenders, we use electrical probes to make contact with on-chip aluminum pads.\\

We apply voltages to the nanobenders in steps of $50$~mV and obtain the reflection spectrum for each voltage (Fig.~\ref{fig:DCtuning}b). We observe wavelength tuning for three different optical modes $\text{O}_i$ of the active cavity. No tuning for the control cavity is observed. Additionally the linear wavelength-voltage relationship around $0$ V indicates that tuning originates from the piezoelectric effect, in contrast to electrostatic, thermo-optical, and thermo-mechanical tuning. Reflection spectra near the fundamental optical mode $\text{O}_1$ around $0$ V and at $2$ V are shown in Fig.~\ref{fig:DCtuning}c. The linewidth of $\text{O}_1$ is around $90$ pm corresponding to a quality factor $Q$ of $1.7\times 10^{4}$. The linewidth is limited by thermal mechanical broadening and decreases by almost an order of magnitude at $4$ K (supplementary information). The shallower dip at $2$ V is due to a decrease of the external coupling rate $\kappa_\text{e}$ as the separation between the zipper cavity and the coupling waveguide is increased by actuation of the nanobender. It may be possible to compensate for this effect by using a secondary nanobender on the coupling waveguide or actuate the two halves of the zipper cavity independently. In Fig.~\ref{fig:DCtuning}d we show the extracted resonance wavelength shift versus DC voltage for $\text{O}_1$ and $\text{O}_2$ of the active zipper cavity. We can tune over tens of nanometers with CMOS-level voltages, corresponding to hundreds of optical linewidths. We perform a linear fit on this tuning curve for small voltages ($|U| < 0.5$ V) and obtain a tuning coefficient quantifying the change in wavelength per volt of $5$~nm/V. All tuning coefficients are reported at $0$ V.\\

We also investigate how tuning coefficient scales with nanobender length (Fig.~\ref{fig:DCtuning}e). For this purpose we fix all other parameters within fabrication imperfections which mostly affect $\tilde{g}_0$. More than $40$ devices with different $L$ are measured. As expected, the zipper cavities with longer nanobenders tune more. The tuning coefficients are higher on devices without the tethers. This is partly supported by simulations. Hence optimizing the way nanobenders are attached is important for composite mechanical systems.

\begin{figure*}[tb]
  \includegraphics[scale=0.9]{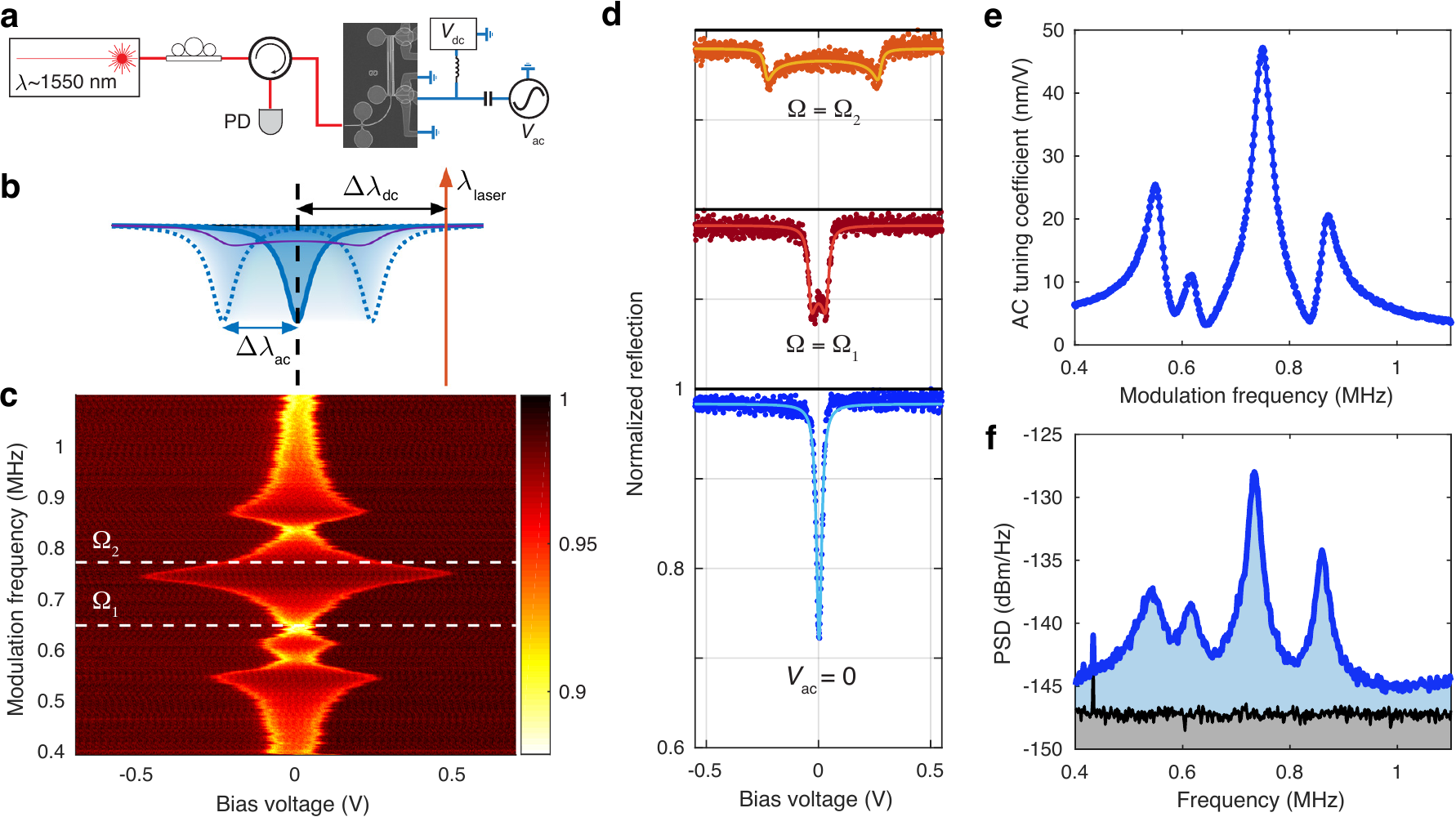} 
\caption{\label{fig:ACtuning}\textbf{AC modulation of a bender-zipper cavity. a}, Experimental setup. \textbf{b}, Measurement principle: as the mode is being modulated at frequency $\Omega$, its resonance wavelength is swept across a laser of fixed wavelength by changing the bias voltage. This results in a measurement of the purple curve. \textbf{c}, Measurement results for a modulation voltage of $50$ mV on the device from Fig.~\ref{fig:DCtuning} showing an enhanced response for certain modulation frequencies. \textbf{d}, Selected cut-lines of Fig.~\ref{fig:ACtuning}c close to resonance, off-resonance and with no modulation. The solid line is an analytical fit of the data. \textbf{e}, AC tuning coefficient as a function of modulation frequency. It is extracted from fitting Fig.~\ref{fig:ACtuning}c and converting bias voltage to wavelength using the DC tuning coefficient. \textbf{f}, Thermal-mechanical power spectral density (PSD) of the bender-zipper cavity. For the blue curve, the laser wavelength is close to the optical resonance, for the black curve it is far detuned.}
\end{figure*}

\textbf{Modulation speed of the bender-zipper cavity.} In addition to slowly tuning the bender-zipper cavity using a DC voltage, we also apply a small AC voltage. This allows us to learn about the AC modulation strength as well as the mechanical resonance frequencies of the bender-zipper device. As shown in Fig.~\ref{fig:ACtuning}a, the total voltage applied on the nanobenders is $V_\text{dc}+V_\text{ac}\sin(\Omega t)$ where $\Omega$ is the modulation frequency. These voltages lead to wavelength shifts of the cavity given by $\Delta\lambda_\text{dc}+\Delta\lambda_\text{ac}(\Omega)\sin(\Omega t + \phi)$ where $\phi$ is a phase offset. In the DC measurements, we sweep the laser wavelength across the resonance of the cavity. For AC measurements, we instead fix the wavelength of the laser and sweep the cavity using the bias voltage (see Fig.~\ref{fig:ACtuning}b), while using the AC voltage to modulate the cavity resonance. The measurement result is the convolution of the cavity's Lorentzian lineshape with the probability distribution that samples the sinusoidally modulated cavity center frequency.%

Sweeping the modulation frequency (Fig.~\ref{fig:ACtuning}c), we observe that the AC tuning coefficient $\alpha_\text{ac} \equiv \Delta\lambda_\text{ac}/V_\text{ac}$ is enhanced at certain frequencies close to $1$ MHz. These correspond to the mechanical resonances of the bender-zipper device (supplementary information). The data was taken with $V_\text{ac} = 50$ mV. Cut-lines of the dataset are shown in Fig.~\ref{fig:ACtuning}d, both off-resonance ($\Omega_1$) and close to resonance ($\Omega_2$). We also show a measurement without AC modulation where we recover a simple Lorentzian. We fit the reflection spectra to extract the AC tuning coefficient and plot it as a function of the modulation frequency (Fig.~\ref{fig:ACtuning}e). Consequently, we are not only able to observe the mechanical resonance frequencies of the system but also directly extract the strength of the modulation. On mechanical resonance, the tuning coefficient is enhanced by a factor $\sim 10$, amounting to $\alpha_\text{ac} \sim 50$ nm/V. This corresponds to $V_\pi = \kappa/(2\alpha_\text{ac}) \sim 1$ mV. 
As expected, the frequency dependence of the AC tuning coefficient closely matches with the thermal-mechanical spectrum (Fig.~\ref{fig:ACtuning}f). We obtain the thermal-mechanical spectrum by detuning the laser from the cavity by around half a linewidth where the cavity frequency fluctuations are transduced to intensity fluctuations that we detect with a high speed detector and record on a real-time spectrum analyzer. The mechanical quality factor $Q_{\text m} \approx 20$ is relatively low due to air damping. This is verified by measurements in low pressure conditions which show several orders of magnitude enhancement in $Q_{\text m}$ (supplementary information). Thus, modulation experiments at low pressures could enable even larger resonant AC tuning coefficients (over a smaller bandwidth), reducing $V_\pi$ to $\sim \SI{20}{\micro\volt}$.

\begin{figure*}[t]
  \includegraphics[scale=0.9]{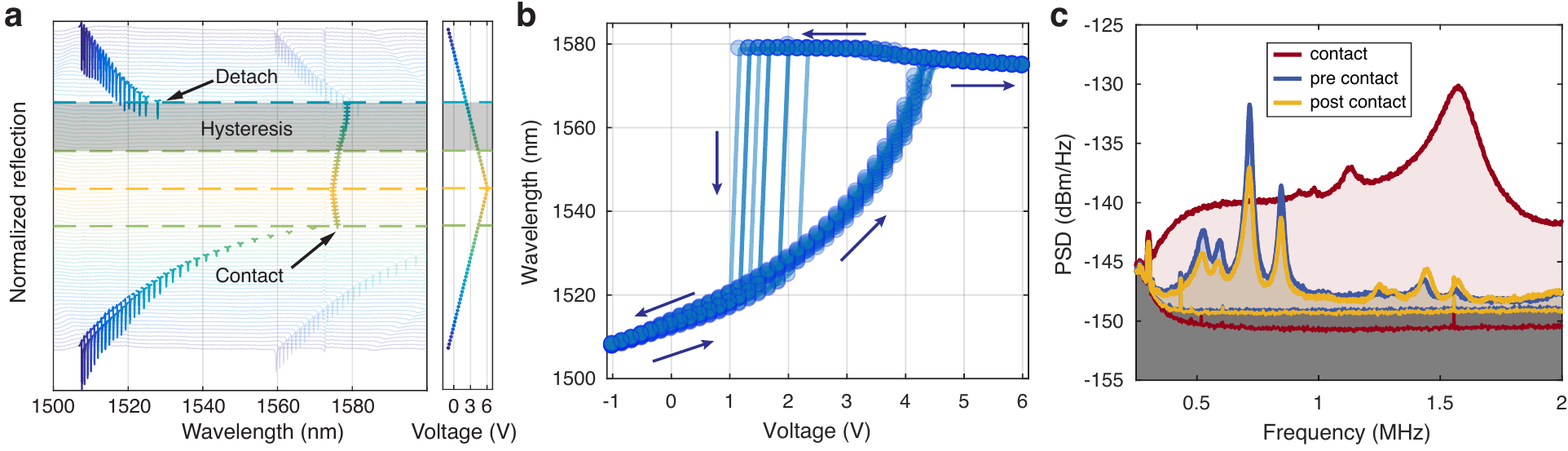}
\caption{\label{fig:contact}\textbf{Mechanical contact of a bender-zipper cavity. a}, One cycle of contact and release of both halves of a zipper cavity attached to $ L= \SI{22}{\micro\meter}$ nanobenders via tethers. As the voltage increases from $-1$ V to $ 4.5$ V in steps of $166$ mV, the resonance wavelength of the zipper cavity tunes up from $1508$ nm to  $ 1576$ nm until both halves of the bender-zipper cavity collide (green dashed line). Further increasing the voltage up to $6$ V (yellow dashed line) only has a small effect. Decreasing the voltage, we observe hysteresis (grey region). The blue dashed line indicates where the zipper detaches. $L=$~\SI{22}{\micro\meter} and $\tilde{g}_0 = 210$ nm (measured on a scanning electron micrograph, pre-release). \textbf{b}, Repeating 9 cycles of contact and release of a zipper cavity revealing a hysteresis loop. \textbf{c}, Thermal power spectral density (PSD) at different stages of a cycle. The mode structure changes when the zippers are making contact.}
\end{figure*}

\textbf{Mechanical contact and hysteresis.} We have shown that tens of millivolts are sufficient to tune the optical cavity by more than its linewidth. The small gap and large displacement per volt, taken together, means that the two halves of the zipper touch for voltages on the order of $5$ V. %

We demonstrate continuous wavelength tuning of a bender-zipper cavity with $L=\SI{22}{\micro\meter} $ by reducing the gap between the two halves of the zipper down to the point when they come into contact (Fig.~\ref{fig:contact}a). Focusing on the fundamental mode $O_1$, we measure a tuning range of $63$ nm with $4.5$ V. To the best of our knowledge, this is the largest tuning range demonstrated for an on-chip optical cavity using CMOS-level voltages. From the initial gap size, we infer a displacement actuation of $\sim 25$ nm/V from each pair of nanobenders. After the contact, the tuning stops regardless of increasing voltage.

As we begin decreasing the voltage, the resonance wavelength shifts ten times less than before the contact because the zipper halves are stuck. We find that the voltage needs to be reduced lower than the contact voltage for the tuning to be restored.  This hysteresis is likely due to the van der Waals force that keeps the zippers attached. When we further decrease the voltage, the nanobenders exert a force opposite in direction which eventually manage to detach the zippers. The whole process is reversible as the mode recovers its original wavelength after detaching.\\

The hysteresis behavior could be applied as an optical memory which necessitates hysteresis for functioning. We test the reliability of the hysteresis loop by repeating the contact-detach process. In Fig.~\ref{fig:contact}b we show nine successive contact-detach cycles, which were preceded by $\sim 40$ cycles. The hysteresis loop is apparent and there is relatively good overlap between the cycles. However the voltage at which the zippers detach is not consistent across cycles and drifts to lower voltages. After several cycles, the nanobenders are not able to get the zippers to detach (not shown here) although we have found that applying a short AC pulse on mechanical resonance is able to detach them reliably, acting as a reset operation. After the reset, for several cycles the zippers are again able to detach with a DC voltage. The reason for this behavior will be subject to future investigations.

 In Fig.~\ref{fig:contact}c we show measurements of the thermal power spectral density of the bender-zipper cavity before contact, during contact and after detaching. We see a clear difference in the spectra between the detached zipper and the attached one. In the latter case, the lateral relative motion between the two halves of the zipper cavity is effectively suppressed. The higher noise floor measured during the contact is likely from laser phase noise, which is more efficiently transduced due to a narrower optical linewidth. We are thus able to reversibly modify both the optical and mechanical properties of the zipper cavity using the nanobenders.

\section*{\label{sec:conclusions}Discussion}

\begin{figure*}[tb]
  \includegraphics[scale=0.9]{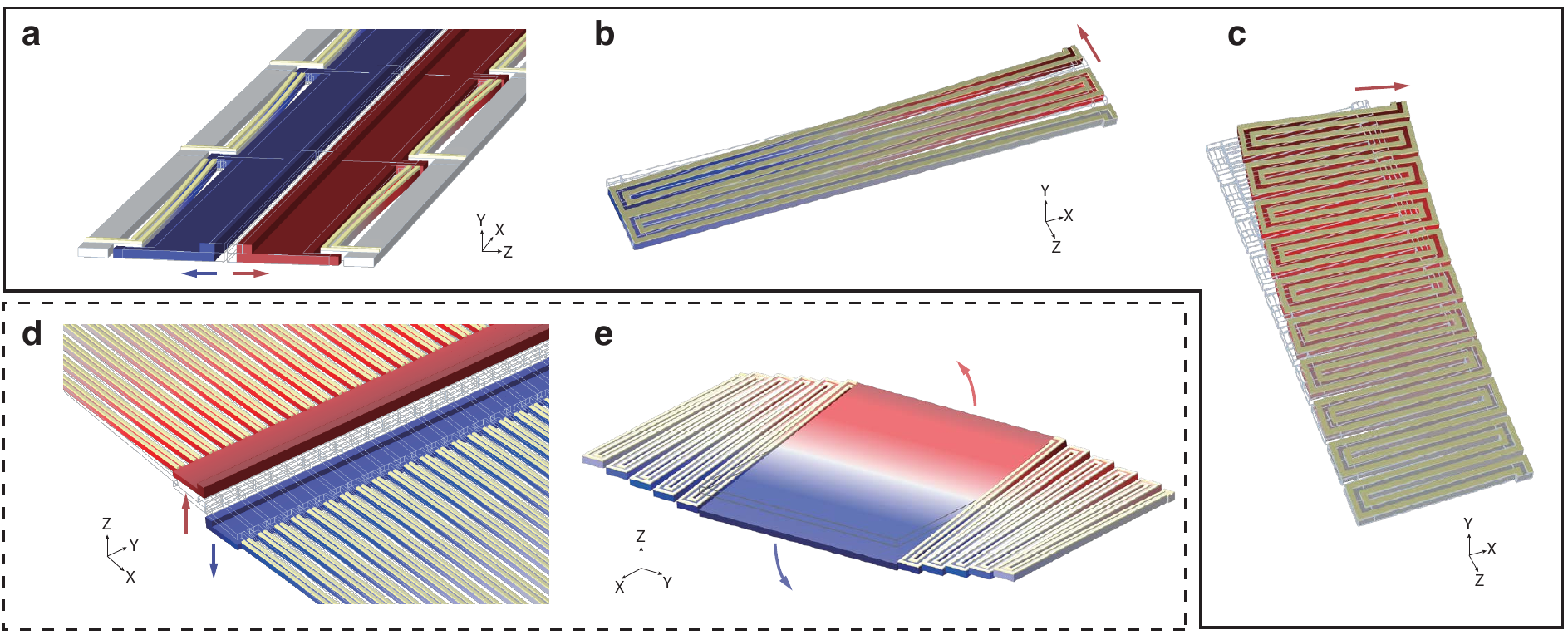}
\caption{\label{fig:outlook}\textbf{Outlook for nanobenders.} Simulated displacement is shown with the metal electrodes colored in bright yellow and lithium niobate colored with relevant displacement field. Directions of the displacements are indicated by colored arrows. The undeformed geometries are outlined in grey. \textbf{a}, Attaching a sliced ridge waveguide to an array of in-plane nanobenders on $Y$-cut LN. The sliced ridge waveguide could operate as either a tunable phase shifter or a tunable coupler. \textbf{b, c,} On nanobenders connected in series in a zig-zag fashion, the displacement accumulates and can be actuated along different directions by engineering the aspect ratio of the zig-zag pattern. \textbf{d}, Nanobenders connected in parallel on $Z$-cut LN. A similar non-uniform $E$ field generates out-of-plane bending. A tunable optical coupler can be realized. \textbf{e}, $Z$-cut nanobenders connected in a zig-zag pattern. An out-of-plane twist accumulates along the zig-zag structure and can be utilized to tilt a nanophotonic mirror or diffraction grating (not shown in detail).}
\end{figure*}

Beyond the demonstrated tunable bender-zipper nanophotonic cavity, various tunable nanophotonic components including phase shifters and couplers can be realized with nanobenders. We show some examples in Fig.~\ref{fig:outlook}. An array of nanobenders connected in parallel to a sliced ridge waveguide acts as a phase shifter. The two halves of the sliced ridge waveguide shift towards or away from each other, effectively tuning the index of the fundamental TE mode~\cite{midolo2018nano,papon2019nanomechanical}.

For applications where a large displacement per volt is desired, a series of nanobenders could be connected in a zig-zag fashion to reduce the length of the occupied region. The direction of the actuated displacement can be controlled by engineering the aspect ratio of the zig-zag pattern (supplementary information). We show one zig-zag bender with four $L=\SI{15}{\micro\meter}$ nanobenders and one with twenty $L=\SI{4}{\micro\meter}$ nanobenders (Fig.~\ref{fig:outlook}b \& c). They have similar simulated displacement per volt of $ \sim \SI{50}{\nano\meter/\volt} $ and actuated along two different directions.

Until now, we have considered nanobenders fabricated with $Y$-cut LN, where the dominant displacement is in-plane, parallel to the surface of the chip. Nanobenders with identical geometry fabricated on $Z$-cut LN and parallel to the crystal $X$ axis, would generate vertical bending (supplementary information). As shown in Fig.~\ref{fig:outlook}(d), these nanobenders can be connected in parallel, for extra structural support, and used to implement a tunable optical coupler~\cite{han2015large,seok2016large}. Moreover, when connected in a zig-zag fashion, a twist is accumulated along the zig-zag. This can be applied to tilt a mirror attached to the end of the zig-zag structure (Fig.~\ref{fig:outlook}e). The rotation angle per volt, actuation speed and device footprint of this type of piezoelectric micro-mirror are comparable to those of the widely used Digital Mirror Devices (DMD)~\cite{dudley2003emerging}.

To summarize, we have introduced and implemented the nanobender, a component capable of generating tens of nanometers of displacement per volt using the piezoelectric effect at the submicron scale. We have experimentally shown tuning of a photonic resonator over the entire telecom C-band with CMOS-level voltages and proposed several new photonic devices that leverage the capabilities of the nanobender. Greater control over photonic and phononic devices on the promising thin-film lithium niobate material platform complements on-going efforts to implement ultra-low-power modulation~\cite{wang2018integrated,zhang2019broadband}, nonlinear nanophotonic circuits~\cite{wang2018ultrahigh,chen2019ultra,lu2019periodically}, quantum nanomechanics~\cite{arrangoiz2018coupling,arrangoiz2019resolving}, and microwave optomechanical transduction~\cite{Jiang2019Lithium,shao2019microwave,dahmani2019piezoelectric,jiang2019efficient}. For emerging quantum technologies that require frequency matching nanophotonic cavities to quantum dots, color centers or rare-earth-doped crystals~\cite{englund2005controlling,faraon2011resonant,zhong2015nanophotonic}, our approach benefits from being able to operate at cryogenic temperatures, while avoiding electric fields, excess carriers, and adsorbed gas molecules, all of which have deleterious effects on the cavity or emitter. Finally, the linear voltage-displacement nature of the piezoelectric effect, the ability to engineer the frequency response and the possibility of dense integration with full electrical access make the nanobenders appealing for sensing~\cite{arlett2011comparative,krause2012high,bagci2014optical,mason2019continuous} and energy harvesting~\cite{wang2006piezoelectric,wang2008towards}.

\section*{Acknowledgements}

The authors would like to thank Agnetta Y. Cleland, E. Alex Wollack, Jeremy D. Witmer, Patricio Arrangoiz-Arriola and Rapha\"{e}l Van Laer for helpful discussions. We thank Chris Rogers for technical support. This work was supported by the David and Lucile Packard Fellowship, the Stanford University Terman Fellowship and by the U.S. government through the National Science Foundation (NSF) (1708734, 1808100), Airforce Office of Scientific Research (AFOSR) (MURI No. FA9550-17-1-0002 led by CUNY). R.N.P. is partly supported by the NSF Graduate Research Fellowships Program (DGE-1656518). Device fabrication was performed at the Stanford Nano Shared Facilities (SNSF) and the Stanford Nanofabrication Facility (SNF). SNSF is supported by the National Science Foundation under award ECCS-1542152.

\section*{Author contributions}

W.J. and A.H.S.-N. conceived the project. W.J. and F.M.M. designed and fabricated the devices. W.J., F.M.M., and T.P.M. developed the fabrication process. F.M.M. and W.J. conducted the measurements with assistance from R.N.P. and C.J.S.. F.M.M. and W.J. wrote the manuscript with input from all authors. A.H.S.-N. supervised the project.

\section*{Competing interests}

A.H.S.-N., W.J., F.M.M and R.N.P. have filed a provisional patent application 62/935953 about the contents of this manuscript. The remaining authors declare no competing interests.

\section*{Methods}

\textbf{Device fabrication.}
We start with $t \approx 500$ nm thin-film LN on a $\sim \SI{500}{\micro\meter} $ thick silicon substrate. Thickness of the LN layer is measured through ellipsometry. The LN is first thinned to $t = 300$ nm through blanket argon ion milling. We then pattern the LN using electron beam lithography (EBL) by coating the sample with HSQ, a negative electron beam resist (Dow Corning, FOx-16). A 10 nm Ti adhesion layer between the LN and HSQ is evaporated prior to the spin coat. The exposure is followed by a development of the HSQ using $25\%$ TMAH and an electron beam hardening step~\cite{yang2006enhancing}. The pattern is then transferred to LN by argon ion milling the LN not covered by HSQ. We proceed with stripping the leftover HSQ with buffered oxide etch and doing an acid clean with diluted hydrofluoric acid to remove re-deposited armophous LN~\cite{hartung2008fabrication}. A second aligned EBL step patterns the liftoff mask for the submicron electrodes on the nanobenders, this time using a positive CSAR resist (Allresist, AR-P 6200.13). Aluminum of thickness $t_\text{m} = 50$ nm is evaporated and lifted off using Remover PG. To pattern the electric probe pads and the wires connecting them to the submicron electrodes, we use photolithography, and subsequent $250$ nm aluminum evaporation and liftoff. Because the probes pads are sitting on the silicon substrate, the two evaporated aluminum layers overlap at the edge of the $300$ nm LN film. The edges of the chip are diced to expose the tapered optical couplers. Finally we do a masked release of the LN using XeF$_2$ which selectively etches the silicon. Sometimes both halves of the zipper cavity are stuck together after release. We notice that using a scanning electron microscope to charge up the structures can get them unstuck.\\

\textbf{Optical characterization.}
All the measurements are done on reflection and a simplified setup is drawn in Fig. \ref{fig:ACtuning}a. A tunable telecom laser (Velocity TLB-6700 and alternatively santec TSL-550) injects light into an optical fiber. With the help of a variable optical attenuator we can control the optical power. Before reaching the tip of a lensed fiber, the light goes through a polarization controller as well as a circulator (port $1 \rightarrow 2$). The lensed fiber is then aligned with the on-chip edge coupler by maximizing the reflection signal from the on-chip reflector. The typical fiber-to-chip coupling efficiency is $\eta \sim 50\%$. The reflected signal goes back through the circulator (port $2 \rightarrow 3$) and is lead to a photodetector (Newport model 1623). By sweeping the laser wavelength, we directly see the modes as dips in the reflection spectrum.
To directly identify the optical modes, we can make use of the optical nonlinearity of LN. For high optical powers ($\sim \SI{200}{\micro\watt}$), we observe second harmonic generation (SHG) happening inside the cavity using a simple optical microscope and a CMOS camera. This is due to the cavity not being resonant around wavelengths of $775$ nm so that light radiates to free space. This can help us identify the optical mode and tell us if it is located in the control or active zipper cavity.
Furthermore, when measuring the thermal-mechanical PSD of the zipper cavity, the wavelength of the laser is slightly detuned from an optical resonance. Instead of going to the photodetector, reflected light is first sent to an erbium-doped fiber amplifier (EDFA) and subsequently to a high-speed photodetector (Newport model 1554-B). We then measure the mechanical spectrum with a real time spectrum analyzer (Rhode \& Schwarz FSW).\\

\textbf{Extracting the AC tuning coefficient.}
The AC voltage is generated using a signal generator (Rigol DG4102) and combined in a bias tee with the DC voltage. It is applied to the on-chip aluminum pads through electrical probes (GGB Industries,  Picoprobe model 40A-GSG) and alternatively, wire bonding.  Considering strong modulation $\Delta\omega_\text{dc}+\Delta\omega_\text{ac}\cos(\Omega t)$ where $\Delta\omega_\text{dc}$ is the detuning from the mode with no modulation and $\Delta\omega_\text{ac}$ the modulation amplitude. This leads to a time-averaged optical reflection signal $|r^2|$ described by: $|r|^2~=~1~-~\frac{c}{2b}\left[\frac{1}{\sqrt{1-(a-ib)^2}}+\frac{1}{\sqrt{1-(a+ib)^2}}\right]$ where $a\equiv\frac{\Delta\omega_\text{dc}}{\Delta\omega_\text{ac}}$, $b\equiv\frac{\kappa}{2\Delta\omega_\text{ac}}$ and $ c\equiv \frac{\kappa\kappa_\text{e}-\kappa_\text{e}^2}{{\Delta\omega_\text{ac}}^2} $. $\Delta\omega_\text{ac} $ and $\kappa\kappa_\text{e}-\kappa_\text{e}^2$ are the two fitting parameters. Qualitatively, we understand the shape of the curve by noticing that the mode spends more time around the extrema of the sinusoidal modulation, hence two peaks form symmetrically with respect to the original cavity resonance. On the other hand the mode spends the least amount of time at the center as this is where the slope of the sine function is largest. Additionally, because the laser is fixed in the measurement, the small but complicated wavelength-dependent background fluctuations of the measurement setup are no longer present which facilitates faithful fitting of the curve to extract the AC tuning coefficient.

\onecolumngrid
\clearpage

\definecolor{commentColorWTJ}{rgb}{0.6,0.2,0.0}
\definecolor{commentColorASN}{rgb}{0.2,0.6,0.0}
\definecolor{commentColorRNP}{rgb}{0.0,0.0,1.0}
\definecolor{commentColorRVL}{rgb}{0.0,0.5,0.5}

\appendix

\begin{table*}
	\caption{Definition of parameters.} \label{tab:parameters}
	\begin{center}
		\begin{tabular}{ccc}
			Parameter & Description & Useful relation \\ 
			\hline
			\hline
			$L$& Length of a nanobender&\\
			$t$& Thickness of a nanobender&\\
			$w$& Width of a nanobender&\\
			$C$  & Bending curvature of a nanobender & \\
			$\phi$ & Central angle of a deformed nanobender & \\
			$ \theta $  & Deflection angle at the end of a deformed nanobender & $\theta = \phi/2 = CL/2$\\
			$\Delta$  & Displacement at the end of a deformed nanobender & $ \Delta = L\theta = CL^2/2 $\\
			$w_\text{m}$& Width of the metal electrodes &\\
			$w_\text{g}$& Gap between the metal electrodes&\\
			$t_\text{m}$& Thickness of the metal electrodes&\\
			$\tilde g$ & Shortest distance between the two zipper beams&\\ 
			$\tilde{g}_0$ & Initial gap size&\\
			$G_\text{OM}$  & Opto-mechanical coupling & $G_\text{OM}=\frac{\partial \omega_0}{\partial z}$ \\
			$\alpha $ & DC tuning coefficient of the bender-zipper cavity& $ \alpha = d\lambda/dV|_{V=0} $\\
			$ \Omega $ & Modulation frequency in the AC tuning measurement& \\
			$V_{\text{dc}}$& Bias voltage in the AC tuning measurement &\\
			$ V_{\text{ac}} $ & Modulation voltage in the AC tuning measurement& \\
			$\Delta_{\text{dc}} $ & Wavelength change from the bias voltage & $ \Delta_{\text{dc}} = \alpha V_{\text{dc}} $ \\
			$ \Delta_{\text{ac}} $ & Wavelength change from the modulation voltage&\\
			$ \alpha_{\text{ac}} $ & AC tuning coefficient & $ \alpha_{\text{ac}} =  \Delta_{\text{ac}}/ V_{\text{ac}} $\\
			\hline
		\end{tabular}
	\end{center}
\end{table*}

\section{Nanobenders: approximated theory, simulations and geometry dependencies}

\subsection{Derivation of the displacement of one nanobender}
\label{SISubsec:derive-single-bender}

Here we derive an approximate theory for the nanobenders. The quadratic scaling law between displacement and length of the nanobender is derived.

\begin{figure*}[ht]
	\includegraphics[scale=1]{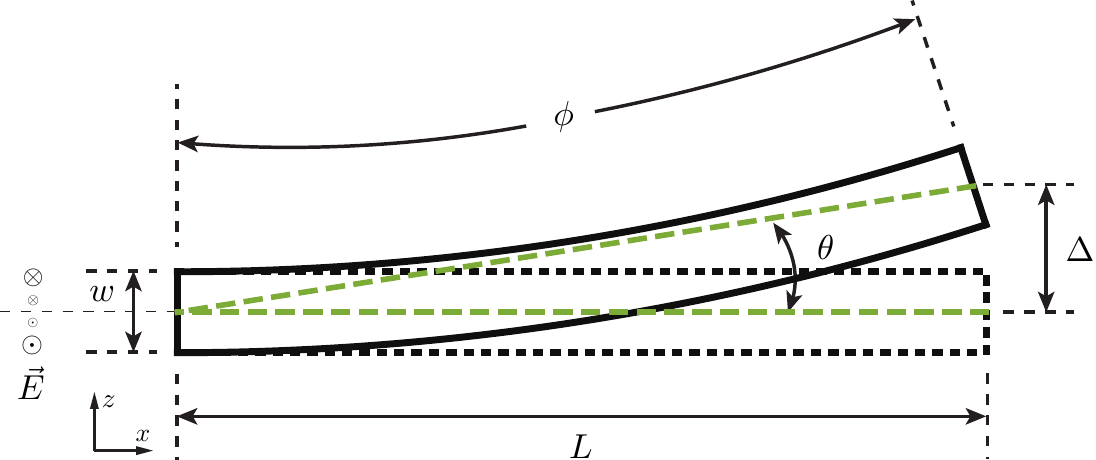} 
	\caption{\label{SIFig:singleBenderDrawing}\textbf{Schematic drawing of a bending beam.} The material coordinate system is chosen and labeled with $x,y,z $. In the case of LN nanobenders, this represents either the side view or the top view of a nanobender, depending on the crystal cut. The direction and spatial variation of the $ \bm E$ field is shown on the left.}
\end{figure*}

Fig.~\ref{SIFig:singleBenderDrawing} shows a single nanobender at equilibrium in dashed black and after deformation in solid black. The left end of the nanobender at $x=0$ is fixed. We tailor the following discussion for application to lithium niobate (LN), thus the coordinate system is chosen to coincide with the material coordinate system of LN. The right end at $x=L$ bends towards the positive $z$ direction by $\Delta$ when a non-uniform electric field $\bm E(\bm r)$ is applied and generates a displacement field $\bm u(\bm r)$. A small angle $\theta$, shown as green dashed lines, is formed by the equilibrium and deformed right end of the nanobender, and the fixed left end. We consider an electric field that is parallel to the $y$ direction (out-of-plane), homogeneous along $x$ and $y$, and varies linearly along the $z$ direction (perpendicular to the nanobender). Consequently, the electric field can be written as $E_y (z) = z \cdot \partial_z E_y$, where $\partial_z E_y $ is spatially homogeneous. We have ignored the effect of a constant electric field which induces a homogeneous strain field, and the generated displacement at the end of the beam is at most proportional to its length $L$.

Bending of the beam can be well described by a radius $R$ and the corresponding curvature $C\equiv 1/R$ when $R\gg L$. The contraction and expansion on the two sides of the beam are related to the bending radius as
\begin{eqnarray}
(R-w/2)\phi &=& L + u_x(x=L, z=w/2),\\
(R+w/2)\phi &=& L + u_x(x=L, z=-w/2),
\end{eqnarray}
where $u_x$ is the $x$ component of the $\bm u$ field, $w$ is the width of the beam and $\phi$ is the corresponding central angle of the deformed beam. Subtracting the two equations, we obtain
\begin{eqnarray}
\phi &=& -\partial_z u_x|_{x=L} = -L\partial_z S_{xx} = -d_{21} L \partial_z E_y,\\
\implies C &=& \phi/L = -d_{21} \partial_z E_y.
\end{eqnarray}

The resulting deflection angle $\theta$ and the deflection $\Delta$ along the $z$ direction are given by
\begin{eqnarray}
\theta &=& \frac \phi 2 = -\frac 1 2  d_{21} L \partial_z E_y = \frac 12 C L,\\
\Delta &\approx& L\theta = -\frac 1 2  d_{21} L^2 \partial_z E_y = \frac 12 C L^2 .\label{SIEqn:Delta-simple-geom}
\end{eqnarray}
We would like to point out that the actual displacement could be either in-plane if $Y$-cut LN is chosen such that the surface of the chip is perpendicular to $y$, or out-of-plane on $Z$-cut LN, where the surface of the chip is perpendicular to $z$, so long as $ \partial_z E_y$ is generated by the electrode configuration. In addition, the above discussion holds for any piezoelectric material with a non-vanishing transverse piezoelectric coefficient.

More generally, the non-uniform electric field can be approximated to first order as
\begin{equation}
E_y(y,z) = y\cdot \partial_y E_y + z \cdot \partial_z E_y.
\end{equation}
The spatially uniform component of the field is ignored. The electrodes for generating the $\bm E$ field are assumed to have a translational symmetry along the direction of the beam ($x$-axis), thus the $E_x$ component is negligible. Furthermore, the magnitude of LN's $d_{21}$ is more than one order of magnitude larger than the other non-zero transverse piezoelectric coefficients. Therefore, we only consider the $d_{21}$ piezoelectric component and the only relevant electric field component is $E_y$. The validity of this approximation is confirmed by simulations in Sec.~\ref{SISubsec:PE-contribution}.

The full piezoelectric constitutive equations in the strain-charge form are 
\begin{eqnarray}
\bm S &=& \bm s_{E} \bm T + \bm d^T \bm E ,\label{SIeqn:straincharge1}\\
\bm D &=& \bm d \bm T + \bm\epsilon_{T} \bm E,
\end{eqnarray}
where $ \bm s_{E} $ is the compliance matrix, $\bm\epsilon_{T}$ is the unclamped permittivity, $\bm D$ is the electric displacement and $\bm T$ is the stress. For a nanobender anchored on one end under one volt applied voltage, the stress in the body of the beam is typically $ \sim 10^6~\text{Pa} $ from simulation. The compliance of LN is $s_{E} \approx 5\times 10^{-12} ~\text{Pa}^{-1}$, thus $ s_E T\sim 5\times 10^{-6} $. On the other hand, the electric field $ E\sim 1~\text{V}/(\SI{200}{\nano\meter}) = 5\times 10^{6} ~\SI{}{\volt \per\meter} $ while $ d\approx 2\times 10^{-11} ~\SI{}{\meter \per\volt} $, leading to $ d\cdot E \sim 10^{-4} $. As a consequence, Eq.~\ref{SIeqn:straincharge1} can be well approximated by $\bm S = \bm d^T\bm E$. This observation agrees with the intuition one would have for a mostly free beam, where the internal stress is expected to be small. Larger stress is localized near the surface of the beam at the boundary of the electrodes, where the electric boundary condition is not continuous.

The displacement field can be solved from the strain field $ S_{xx}(y,z) = d_{21} E_y (y,z)$ and $ S_{ij} = 0 $ for all other components. For a free beam with $\bm u(0,0,0) = (0,0,0) $ and ignoring any rigid rotation,
\begin{eqnarray}
\label{SIEqn:analytic-u-of-x-y-z}
\bm u (x,y,z) &=&  d_{21} \left(x(y \partial_y E_y + z\partial_z E_y), - \frac{1}{2}x^2 \partial_y E_y  , - \frac{1}{2}x^2 \partial_z E_y \right),\\
\Delta &\equiv& u_z(x=L) = -\frac12 d_{21} L^2\partial_z E_y = \frac 12 CL^2. 
\end{eqnarray}
As a result, bending of the beam is observed for both $y$ and $z$ direction, where the displacement at $x=L$ scales quadratically with the length of the beam $L$. The displacement along the $z$ direction is identical to Eq.~\ref{SIEqn:Delta-simple-geom}.

In the above derivation, $ \partial_y E_y$ and $\partial_z E_y $ are assumed to be uniform on the cross-section of the beam. In reality for a $Y$-cut nanobender where a parallel pair of electrodes is placed on the top surface ($x$-$z$ plane) of the beam, the fringing field has a complicated spatial variation. Nevertheless, the field $E_y$ mostly varies along the $z$ direction and the variation along $y$ is negligible after averaging over the cross section. Similarly, the field $E_z$ mostly varies along the $y$ direction. We verify this by numerically evaluating the average of $ \partial_i E_j  $ on the cross section for $i,j = y,z$ using COMSOL simulations. The averaged $ \partial_y E_y $ and $\partial_z E_z $ are two orders of magnitude smaller than $ \partial_y E_z$ and $\partial_z E_y$. This can be qualitatively understood by considering the symmetry of the electrostatic potential and the electric field. The parallel pair of electrodes possess $z$-reflection symmetry, where the geometry of the electrodes is invariant under reflection with respect to the symmetry plane $z=0 $. When a voltage difference is applied to the electrodes and by setting the potential reference such that $V=0 $ on $z=0$, we have approximately $V(y, -z) \approx -V(y, z)$. Consequently,
\begin{eqnarray}
\partial_y E_z(y, -z) & \approx &  \partial_y E_z(y, z), \\
\partial_z E_y(y, -z) &  \approx &  \partial_z E_y(y, z), \\
\partial_y E_y(y, -z) & \approx &  -\partial_y E_y(y, z), \\
\partial_z E_z(y, -z) & \approx &  -\partial_z E_z(y, z).
\end{eqnarray}
The implication from these observations is that $ \partial_y E_y $ and $ \partial_z E_z $ average to roughly zero on the cross section. Intuitively speaking, by virtually dividing the beam into two halves separated by the $z$-symmetry plane, the two halves of the beam tend to bend against each other when $ \partial_y E_y $ and $ \partial_z E_z $ are considered, thus cancelling and leading to an insignificant deformation of the full beam. In contrast, they bend in accordance with each other from $\partial_y E_z $ or $\partial_z E_y$, generating a much larger bending of the full beam. Similarly for a $Z$-cut nanobender where the electrodes are fabricated on the $x $-$y$ plane, $y$-reflection symmetry can be considered, and the same arguments still hold.

As discussed above, the bending from Eq.~\ref{SIEqn:analytic-u-of-x-y-z} is mostly along a single direction $z$, generating a ``clean'' displacement $\Delta\equiv u_z(x=L) $. In other words, for a nanobender where only $d_{21}$ is considered as non-zero and the cross-section is symmetric under $y$ or $z$ reflection, the bending generates a displacement along the $z$ direction.

For example, on a $Y(Z)$-cut LN nanobender, the displacement is mostly in-plane (out-of-plane) where displacements along the other directions are more than one order of magnitude smaller (see Sec.~\ref{SISec:nanobender-sim} for simulation results). The largest contribution to displacements along the other directions is bending generated by a different transverse piezoelectric coefficient.

Lastly we would like to point out that the transverse piezoelectric coefficients ($d_{13} $-type) are not the only components that could potentially generate bending from a non-uniform electric field. Through numerical simulation (Sec.~\ref{SISec:nanobender-sim}), we found that the diagonal components ($d_{11} $-type) and the shear components ($d_{14}$-type) are also capable of generating bending under the parallel-pair electrode configuration.

\subsection{Simulation of one-volt displacement}
\label{SISec:nanobender-sim}

We use the Piezoelectric Devices module in COMSOL for simulating the 3D model of the nanobender. A rotated coordinate system is adopted to account for different crystal cuts and nanobender orientations. A fixed constraint is applied to one end of the nanobender while the other surfaces are subject to a free boundary condition. The metal surfaces are selected and assigned as either ground or a voltage terminal with $V_0 = \SI{1}{\volt}$. The full geometry is encapsulated in air for the electrostatics simulation. We found that the air encapsulation causes only a minor effect due to LN's large relative permittivity. A stationary study is solved for the static displacement, and an eigenfrequency study is solved for the eigenmode frequencies of the nanobender.

For generality and simplicity, from now on we adopt a global coordinate system, with axes denoted by $x,y$ and $z$ that are fixed with the nanobender, where $x$ is parallel to the nanobender and is pointing from the fixed-end to the free-end. Positive $z$ axis is perpendicular to the chip. We report the simulated displacements in this global coordinate system. As a result, for nanobenders parallel to crystal $X$ axis as discussed above, the bending is always towards the crystal $Z$ axis. The cut of the crystal is defined as the crystal axis perpendicular to the plane of the thin film from which the beam is fashioned. Since we often put electrodes on top of the thin film, the example in the beginning of the supplementary information would naturally correspond to a $Y$-cut film where the beam is patterned to be parallel to the crystal $X$. In the newly defined chip coordinate system, we have $z||Y$, $x||X$, $y||Z$, and therefore a large displacement being generated in $y (Z)$ which is in the plane of the chip. Similarly, for $Z$-cut nanobenders, a vertical bending results and leads to a large $u_z$.

\subsubsection{Contributing piezoelectric components to the displacement}
\label{SISubsec:PE-contribution}

To show that the dominant contribution of the displacement is from the $d_{21}$ component of the piezoelectric tensor, we conduct finite element simulations with the original piezoelectric tensor $\bm d$ and compare the results to simulations where all components of $\bm d$ other than $ d_{21} $ are set to zero.

\begin{figure*}[ht]
	\includegraphics[scale=0.9]{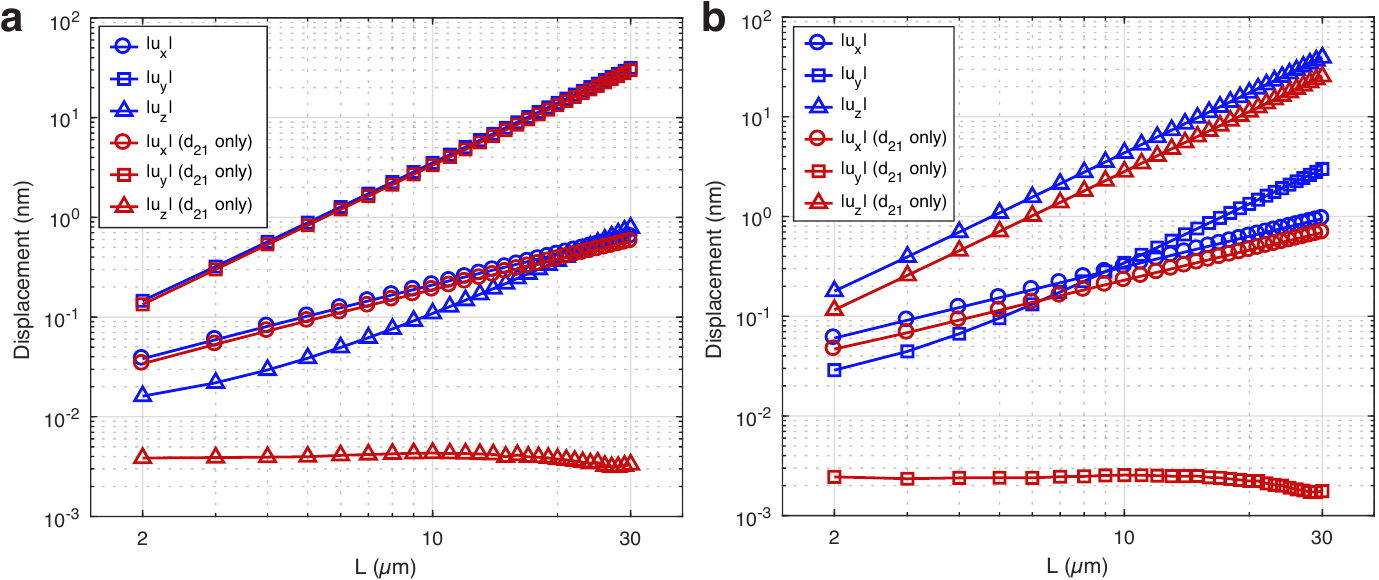} 
	\caption{\label{SIFig:d21}\textbf{Simulated displacements} for $Y$-cut (\textbf{a}) (corresponding to Fig~\ref{SIFig:singleBenderDrawing} with $\Delta=u_y$) and $Z$-cut (\textbf{b}) nanobenders ($\Delta = u_z $). To show the contribution from the $d_{21}$ piezoelectric coefficient, identical geometry is simulated with full $\bm d$ tensor (blue) as well as a $\bm d$ tensor where $d_{21}$ is the only non-zero component (red).}
\end{figure*}

The simulated displacements are plotted in Fig.~\ref{SIFig:d21} for different nanobender lengths $L$ in log scale. For $Y$-cut nanobenders (Fig.~\ref{SIFig:d21}a), the $d_{21}$ component is able to produce the simulated displacement from the full $ \bm d$ tensor with minor deviation. The quadratic scaling of $ u_y $ versus $ L $ can be clearly observed by comparing to $ u_x $. The large deviation between $d_{21}$ only and full $\bm d $ simulations for $u_z$ can be explained by the non-zero $d_{31}$ component of LN, which generates out-of-plane bending from $\partial_y E_z $. Since the magnitude of $ d_{31} $ is more than one order of magnitude smaller than $d_{21}$, the vertical bending is negligible comparing to in-plane bending. However, the quadratic scaling is still present and makes $u_z$ larger than $u_x$ for sufficiently large $L$.

As for vertical nanobenders on $Z$-cut LN, we observe (Fig.~\ref{SIFig:d21}b) that keeping only the $d_{21}$ coefficient underestimates the full simulated displacement. The other piezoelectric components contribute to roughly $ 35\% $ of the displacement $ u_z $. Surprisingly we find that $ 29\% $  out of the $35\%$ displacement that cannot be explained by the $d_{21}$ component originates from the $ d_{22} $ and $ d_{24} $ components. These components are not expected to generate bending from the intuitive explanation provided in Sec.~\ref{SISubsec:derive-single-bender}, and likely result from the complicated non-uniform electric field profile.

\subsubsection{Dependency of displacement on nanobender geometry}
\begin{figure*}[ht]
	\includegraphics[scale=0.9]{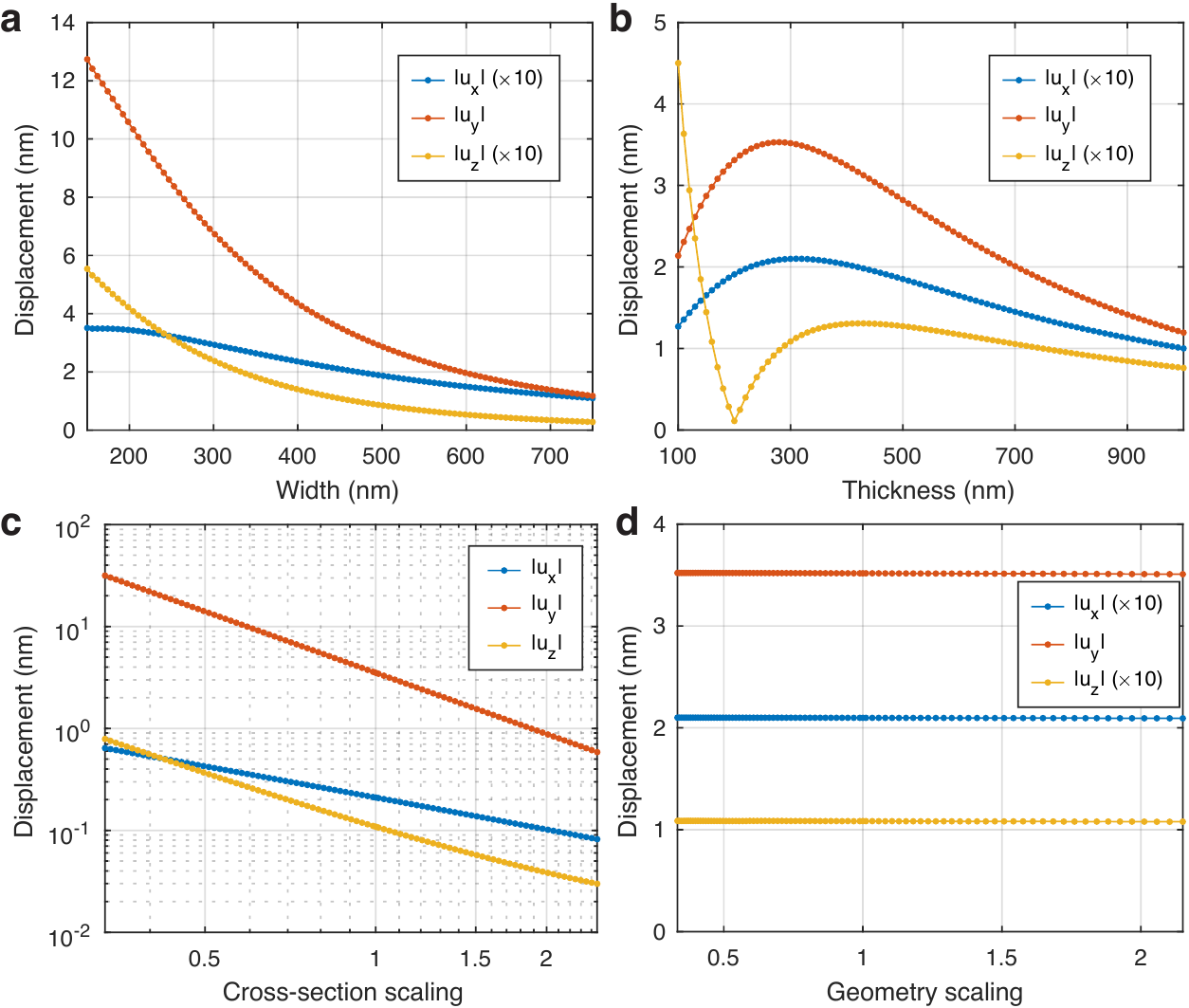} 
	\caption{\label{SIFig:bender_in_plane}\textbf{Simulated one-volt displacement for $Y$-cut nanobenders of varying geometries. a}, Sweep of the width $w$. \textbf{b}, Sweep of the thickness $t$. \textbf{c}, Scaling of the cross-section. \textbf{d}, Scaling of the whole geometry.}
\end{figure*}

So far, we have mostly considered the quadratic length-displacement relationship of the nanobender. Here, we investigate how other geometric parameters affect the displacement for both $Y$-cut and $Z$-cut nanobenders. Figures~\ref{SIFig:bender_in_plane}a and b show how the maximal displacement of the nanobender for all three directions scales with width $w$ and thickness $t$. As expected, the displacement increases for smaller width. We also observe that for varying thickness, the in-plane displacement $u_y$ peaks at around $t=300$ nm for a fixed electrode thickness $t_\text{m}=50$ nm. Furthermore, it is interesting to see how scaling the cross-section for a fixed length $L=$\SI{10}{\micro\meter} affects displacement. Scaling the cross-section relative to $(w,t,t_\text{m})=(450,300,50)$ nm, we observe (fig.~\ref{SIFig:bender_in_plane}c) that as the cross-section increases, $u_y$ and $u_z$ decrease quadratically whereas $u_x$ decreases linearly. Finally, following the scale-invariance argument in the main text, the displacement of the nanobender is not dependent on its relative size. By scaling the whole geometry of the nanobender relative to $(w,t,t_\text{m},L)=(450,300,50,10^4)$ nm, we confirm through simulations that the displacement is indeed independent of geometry scaling.

\begin{figure*}[ht]
	\includegraphics[scale=0.92]{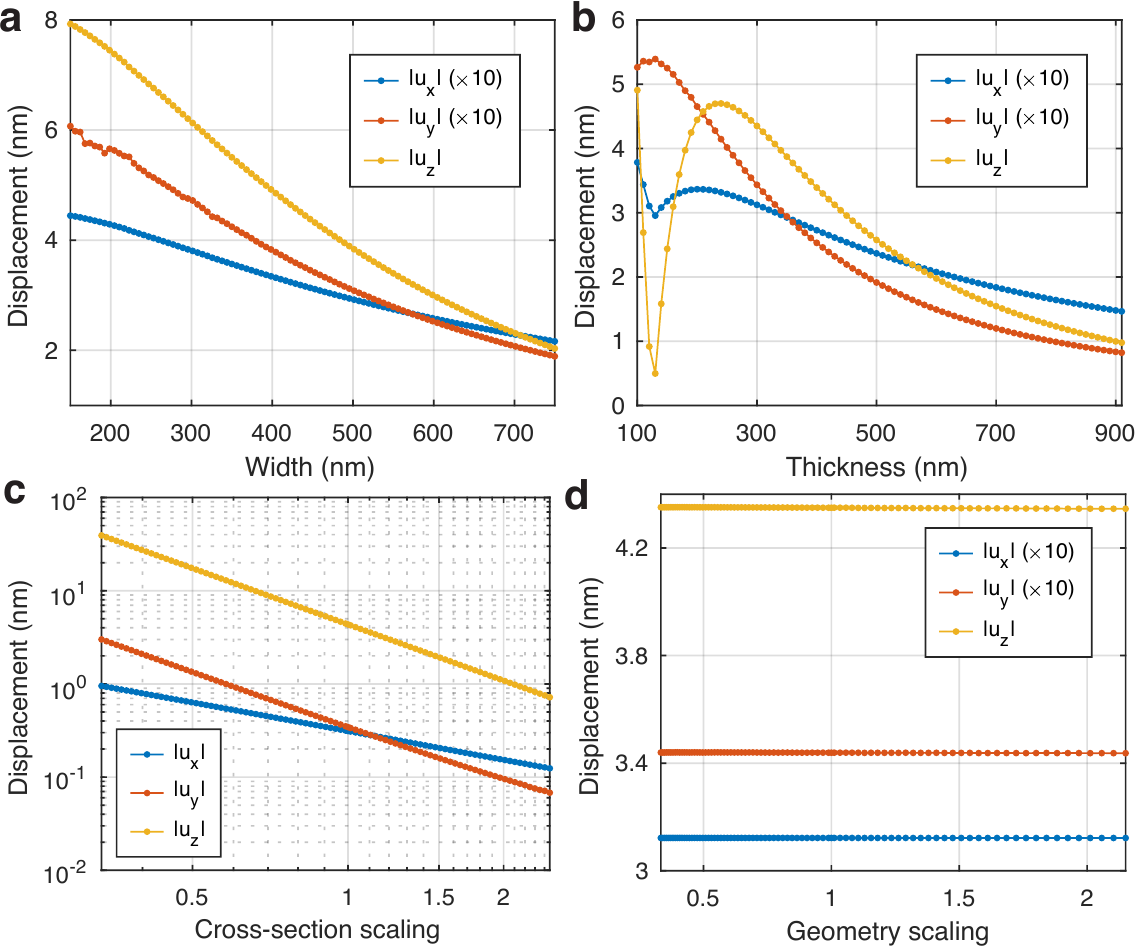} 
	\caption{\label{SIFig:bender_vertical}\textbf{Simulated one-volt displacement for $Z$-cut nanobenders of varying geometries. a}, Sweep of the width $w$. \textbf{b}, Sweep of the thickness $t$. \textbf{c}, Scaling of the cross-section. \textbf{d}, Scaling of the whole geometry.}
\end{figure*}

For $Z$-cut nanobenders (Fig.~\ref{SIFig:bender_vertical}), the simulations and analysis are similar except that the out-of-plane displacement component $u_z$ is dominant instead of $u_y$.

\subsection{Approximated displacement for zigzag nanobenders}

In this section, we show that the nanobender on $Y$-cut LN can be wrapped in a zigzag structure without metal crossover while maintaining efficient displacement actuation. The direction of the in-plane displacement can be controlled by the aspect ratio of the zigzag structure. Furthermore, the identical zigzag configuration generates out-of-plane deflection that accumulates along the zigzag on $Z$-cut LN.

\begin{figure*}[ht]
	\includegraphics[scale=0.9]{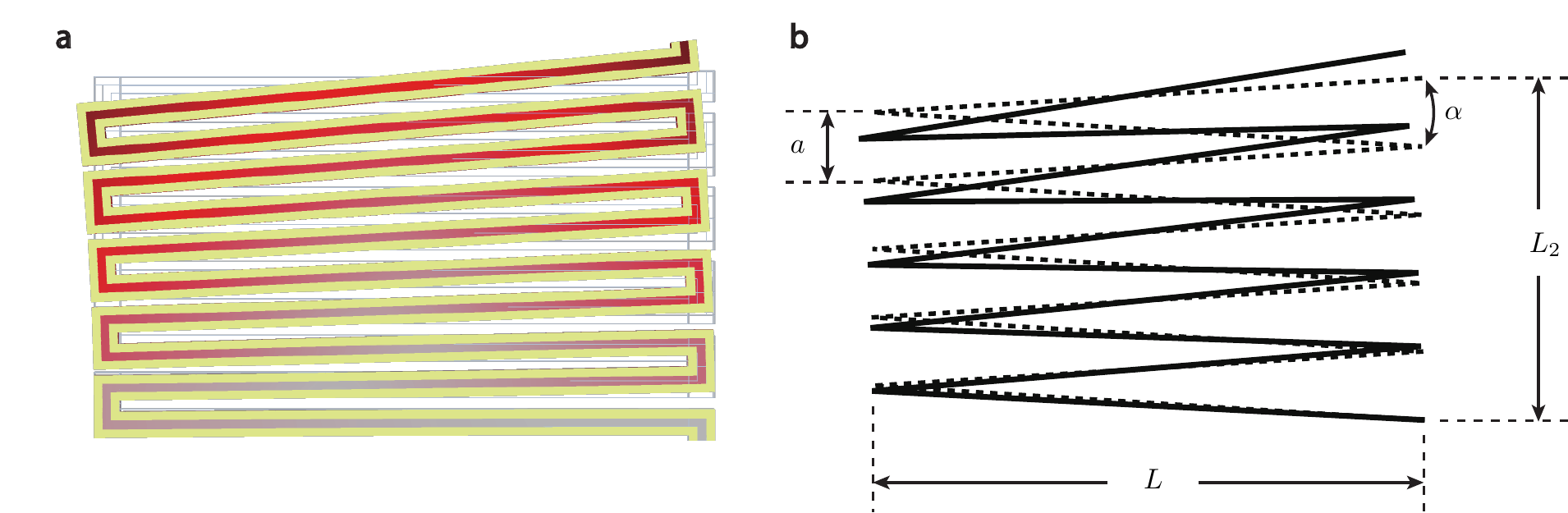} 
	\caption{\label{SIFig:zigzag_Y}\textbf{Zigzag nanobenders on $Y$-cut LN for in-plane displacement.} \textbf{a}, Simulated displacement for a zigzag nanobender. The bottom-right corner of the zigzag structure is fixed. The LN domain is colored by displacement and the electrodes are colored in yellow. \textbf{b}, Simplified geometry of the zigzag nanobender. The geometry at equilibrium is shown as dashed lines, and the deformed geometry is shown as solid lines. }
\end{figure*}

We start with a $Y$-cut zigzag nanobender. Fig.~\ref{SIFig:zigzag_Y} shows a simulated displacement profile (Fig.~\ref{SIFig:zigzag_Y}a) and a schematic drawing of the simplified zigzag geometry (Fig.~\ref{SIFig:zigzag_Y}b). We simplify one unitcell of the zigzag structure as two solid lines which connect at the ``U-turns'' and form an angle $\alpha = 2\arctan\left(a/2L\right)\approx a/L $. $L$ is the length of a single nanobender and $a$ is the width of one unitcell. The extension of the zigzag structure is $L_2\equiv Na$ where $N$ is the number of unit cells.

Since only the in-plane displacements are relevant, the coordinates and displacements can be represented by complex numbers. We establish a complex plane such that the real axis is parallel to a single nanobender and the origin coincides with the bottom right of the zigzag structure where it is anchored. To move along the vertices of the simplified zigzag, two complex numbers $ v_1 = -L+i a/2 $ and $v_2 = L + i a /2 $ can be added consecutively. As a result, for a zigzag nanobender with $N$ unit cells ($2N$ connected nanobenders), the position of the end point $x_N$ is
\begin{equation}
x_N = N(v_1+v_2) = iNa.
\end{equation}

When a voltage is applied such that a single nanobender is deflected by an angle $\theta$ as defined in Sec.~\ref{SISubsec:derive-single-bender}, it will be shown in the following that the displacement at $x_N$ is solely determined by $\theta, L $ and $L_2$.

As shown in Fig.~\ref{SIFig:singleBenderDrawing}, every nanobender in the zigzag structure contributes a rotation of $ \exp(i\phi) $ from one vertex to the next vertex. Note that the rotation is $\phi = 2\theta$ instead of $\theta$. One would expect $\theta$ if only considering the displacement. However, the next nanobender is connected tangentially to the end of the previous nanobender, which, due to the bending, has changed its orientation by $\phi$. The end position $x'_N$ of the zigzag after the series of bending and rotation can be expressed as
\begin{eqnarray}
x'_N &=& v_1 e^{i\phi} + v_2 e^{2i\phi} + \cdots + v_1 e^{(2N-1)i\phi} + v_2 e^{2Ni\phi} \notag\\
&=&\left( v_1 e^{i\phi} + v_2 e^{2i\phi}  \right) \sum_{k=0}^{N-1} \exp(2ik\phi) \notag\\
&=&\left( v_1 + v_2 e^{i\phi} \right) \frac{ \sin(N\phi) }{\sin \phi} e^{iN\phi}.
\end{eqnarray}
The resulting displacement at the end of the zigzag is
\begin{eqnarray}
u_N &\equiv & x'_N - x_N \approx \left( v_1 + v_2 (1 + i\phi) \right)\cdot N\cdot (1+iN\phi) - (v_1+v_2)\cdot N \notag\\
&\approx& (v_1 + v_2)\cdot N \cdot iN\phi +  i\phi v_2 N \notag\\
&=& N\phi \cdot ( -aN + i v_2 ) \approx N\phi (-L_2 + iL).
\end{eqnarray}
We have expanded the expression up to first order in $\phi $, used the relationship $ v_1 + v_2 = ia $ and the approximation $ v_2 = L + ia/2 \approx L $. Written back in vector notation, we obtain the in-plane displacement
\begin{equation}
\label{SIEqn:Y-cut-zigzag-esti}
\bm{u}_N = N\phi ( -L_2, L ) = CNL (-L_2, L),
\end{equation}
where $C $ is the bending curvature that can be obtained by a single nanobender simulation, which is much faster than simulating the full zigzag structure. The total displacement is enhanced by the number of unit cells $N$, and the direction of the displacement is related to the overall geometry of the zigzag structure $ L $ and $L_2$.

To compare the above simplified estimation to the simulated displacement, we evaluated the maximal displacements from three different zigzag structures using COMSOL and using the estimation Eq.~\ref{SIEqn:Y-cut-zigzag-esti}. The comparison is shown in Table \ref{tab:in-plane-zigzag-comparison}. A one-volt curvature of $ C = \SI{95.2 }{\per\meter} $ obtained from a single nanobender simulation is used. We see reasonable agreement between simulated and estimated displacements for zigzags with various aspect ratios $ L/L_2 $, where the deviations are all $ \lesssim 20\% $.

\begin{table*}
	\caption{\label{tab:in-plane-zigzag-comparison}Simulated and estimated one-volt displacement for $Y$-cut zigzag nanobenders.}
	\begin{center}
		\begin{tabular}{ccc|cc}
			$N$ & $L$ (\SI{}{\micro\meter}) & $L_2$ (\SI{}{\micro\meter}) & Simulated $ \bm u $  (\SI{}{\nano\meter}) & Estimated $\bm u$ (\SI{}{\nano\meter}) \\ 
			\hline
			$10$ & $4$ & $13$ & $ (53.4, 20.6,0.14) $ & $ (49.5, 15.2, 0) $\\
			$2$ & $15 $ & $2.6$ & $ (7.3, 46.8, 0.37) $ & $ (7.4, 42.8, 0) $\\
			$5$ & $ 10 $ & $6.5 $ & $ (31.6, 54.1, 0.30) $ & $ (30.9, 47.6, 0) $
		\end{tabular}
	\end{center}
\end{table*}

\begin{figure*}[ht]
	\includegraphics[scale=0.9]{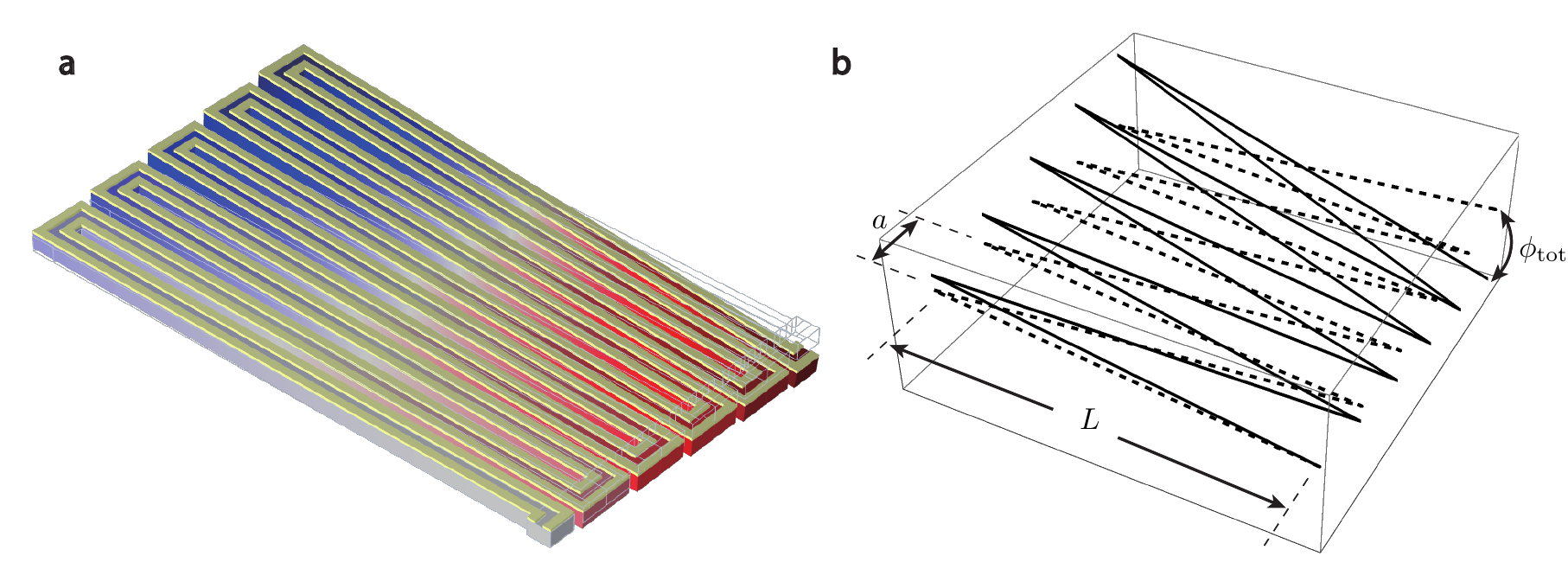} 
	\caption{\label{SIFig:zigzag_Z}\textbf{Zigzag nanobenders on $Z$-cut LN for out-of-plane deflection.} \textbf{a}, Simulated displacement for a zigzag nanobender. The bottom-right corner of the zigzag structure is fixed. The LN domain is colored by the out-of-plane displacement and the electrodes are colored in yellow. \textbf{b}, Simplified geometry of the zigzag nanobender. The geometry at equilibrium is shown as dashed lines, and the deformed geometry is shown as solid lines. }
\end{figure*}

When $Z$-cut zigzag nanobenders are considered for out-of-plane actuation, the situation is simpler comparing to the $Y$-cut in-plane zigzag nanobenders. The direction of bending is disentangled with the geometry of the zigzag structure, and the rotation simply accumulates along the zigzag structure (Fig.~\ref{SIFig:zigzag_Z}).

More specifically, at the end of each unit cell of the deformed zigzag, the tangential direction is changed by $ 2\phi $. After $N$ unit cells, the accumulated angle is $ \phi_{\text{tot}} = 2N\phi = 2NCL $ where $C $ is the bending curvature. For a one-volt $ C \sim \SI{100}{\per\meter} $, $ L = \SI{20}{\micro\meter}$ and $N = 10$,  $ \phi_{\text{tot}} \sim 0.04 = 2.3$ degrees. A deflection of $ \pm 23 $ degrees can be achieved with voltages between $ \pm \SI{10}{\volt} $.

As for the corresponding vertical displacement, it is directly given by $ \Delta_{\text{zz}} \approx \phi_{\text{tot}}L/2 = NCL^2 $. This is much smaller than the displacement generated by a single nanobender with the same total length $L_{\text{tot}} = 2NL $, which is $ \Delta \approx CL_{\text{tot}}^2/2 = 2N^2CL^2 $. Since a large displacement is transduced, one expects the effective stiffness of the single nanobender with length $ L_{\text{tot}}$ to be much smaller than the zigzag structure. To verify this, we simulated the displacement of a zigzag nanobender with $L=\SI{10}{\micro\meter}$ and $N=5$ under a vertical force of $ f_v = \SI{1}{\nano\newton} $ at the end of the zigzag, and also the displacement of a single nanobender with $ L_{\text{tot}} = \SI{100}{\micro\meter}$ under the same conditions. The simulated displacements are $ \Delta_{\text{zz}} = \SI{27.6}{\nano\meter} $ for the zigzag and $ \Delta = \SI{1.5}{\micro\meter} $ for the single nanobender.

From Euler-Bernoulli beam theory (see also Sec.~\ref{SIsubsec:mech-freq}), the effective spring constant of a beam scales as $k\propto L^{-3}$. By approximating the joint between every nanobenders in the zigzag structure as a stiff connection, the effective spring constant of the zigzag is $ k_{\text{zz}} \propto L^{-3}/(2N)$, where the proportionality coefficient is determined by material properties and the cross section geometry. On the other hand, the single nanobender has $ k \propto L_{\text{tot}}^{-3} = L^{-3}/(2N)^3 $. As a result, the single nanobender is a factor of $ 4N^2 $ less stiff than the zigzag nanobender. Note that the central angle generated by the single nanobender is $ \phi = CL_{\text{tot}} = 2NCL $, identical to the zigzag nanobender. Hence for applications where an out-of-plane rotation is desired, the zigzag nanobender is advantageous over a single nanobender, where the zigzag nanobender generates an identical out-of-plane rotation with a much stiffer structure.

\section{Nanophotonic zipper cavity design and properties}

\subsection{Mirror cell design}
\begin{figure*}[ht]
	\includegraphics[scale=0.95]{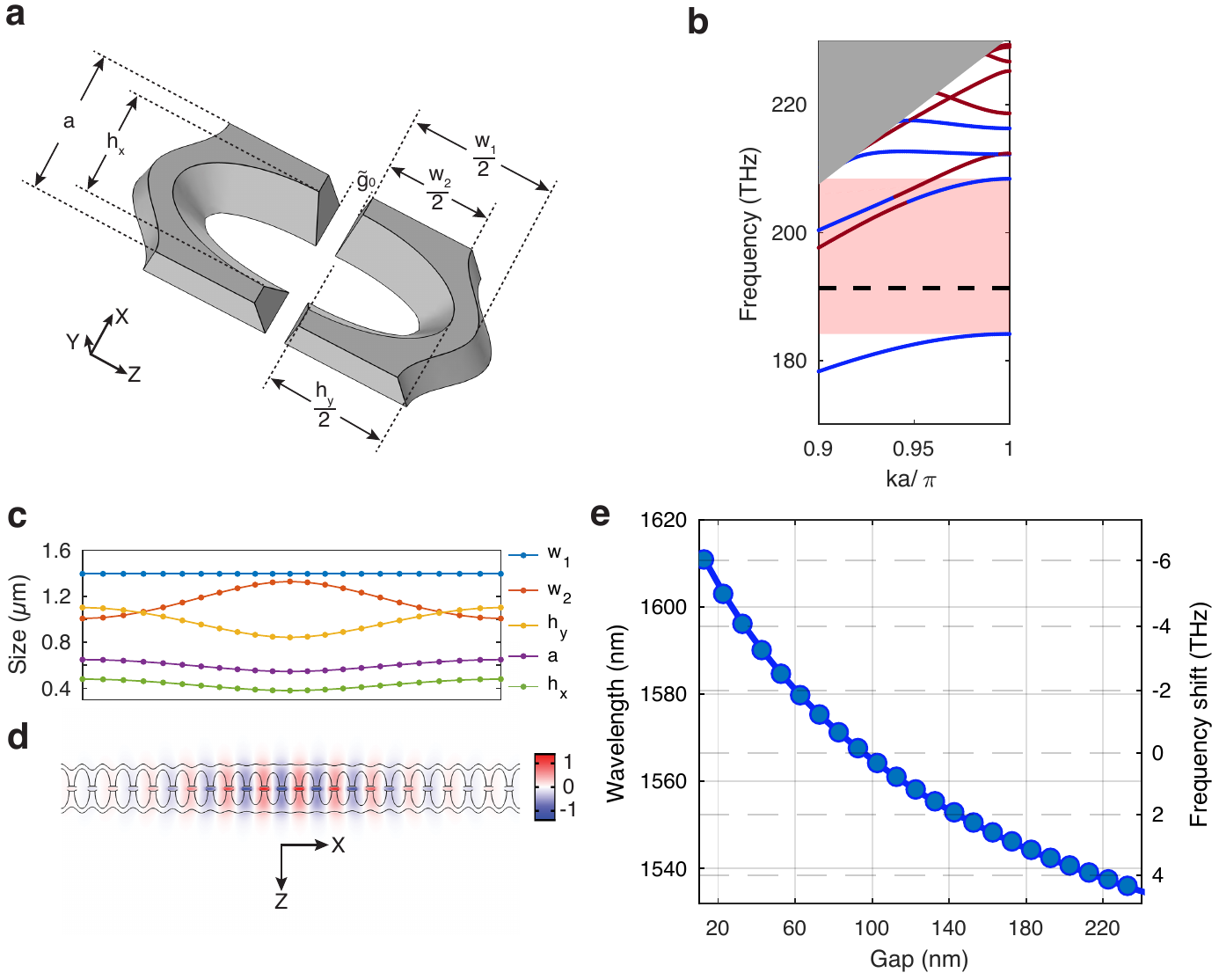} 
	\caption{\label{SIFig:zipper}\textbf{Nanophotonic zipper cavity design and properties. a}, Unit cell geometry of the LN zipper cavity. \textbf{b}, Optical band diagram of the unit cell geometry. The dashed line indicates the localized fundamental optical TE mode. The grey region corresponds to the radiation continuum. \textbf{c}, Zipper cavity unit cell parameters along the transition from mirror cell to defect cell and back. \textbf{d}, Normalized $E_Z$ component of the fundamental optical mode of the zipper cavity. \textbf{e}, Resonance wavelength versus zipper gap $\tilde{g}_0$.}
\end{figure*}

In the following, we briefly describe the procedure, adapted from Ref.~\cite{Jiang2019Lithium}, for designing the $Y$-cut lithium niobate mirror cell used as part of the 1D zipper nanophotonic cavity and the reflector at the end of the coupling waveguide. The thickness of the mirror cell is set to $t=300$ nm. Because of fabrication imperfections, the LN structures have angled sidewalls which we take into account in the design. We therefore use an outer angle (defined as the angle between the sidewall and the vertical direction) of $11^\circ$ and inner angle (inside the hole) of $22^\circ$, based on scanning electron micrographs. The mirror cell geometry (Fig.~\ref{SIFig:zipper}a) consists of two identical halves separated by a gap $\tilde{g}_0$. Each half is constructed by adding a cosine edge with amplitude $(w_2-w_1)/4$ to a rectangle with dimensions $a$ by $(w_1+w_2)/2$. Half of an ellipse (with axes $h_x$ and $h_y$) is then cut out of the rectangle to generate an air-hole. By periodically continuing the unit cell with lattice spacing $a$, we can open a quasi-TE optical bandgap at the $X$-point (see Fig.~\ref{SIFig:zipper}b). Based on eigenfrequency simulations and optimization of the bandgap size, we choose the parameters $(a,w_1,w_2,h_x,h_y)=(650,1395,1007,481,1103)$ nm leading to a quasi-bandgap of $24.3$ THz centered at $196.3$ THz. We used multivariable genetic optimization to determine the parameters for this geometry. The cost function tries to maximize the size of the quasi-TE bandgap at the $X$-point while trying to push up the fundamental quasi-TM mode out of the bandgap to avoid scattering into it. The gap $\tilde{g}_0$ is set to $94$ nm.

\subsection{1D nanophotonic cavity}

To localize an optical mode inside a 1D nanophotonic cavity, it is standard to introduce a defect unit cell inside an array of mirror cells with a smooth set of transition cells in between. This defect cell has a fundamental quasi-TE mode at the $X$-point that is inside the bandgap of the mirror cell. To generate the defect cell starting from the mirror cell, it is thus necessary to push up the fundamental quasi-TE mode of the mirror cell. This is achieved by reducing its lattice spacing $a$ as well as the relative air-hole size. The defect and mirror cells are smoothly connected through cubic interpolation \cite{chan2012laser} using $12$ transition cells (see Fig.~\ref{SIFig:zipper}c). We additionally have $8$ mirror cells on each side of the cavity. 

We determine the geometry of the defect cell with a similar multivariable genetic optimization. The cost function maximizes the optical quality factor $Q$ of the localized fundamental optical mode while disregarding solutions that lead to unphysically high $Q$s~\cite{Jiang2019Lithium}. As a result of this optimization, a defect cell geometry with parameters $(a,w_1,w_2,h_x,h_y)=(547,1395,1329,381,843)$ nm is chosen, and the number of transition cells and mirror cells are determined to be $12$ and $8$ respectively. Fig.~\ref{SIFig:zipper}d shows the simulated fundamental quasi-TE mode profile for this design choice. Furthermore, we see that the resonance wavelength of the cavity is strongly dependent on the gap $\tilde{g}_0$ of the cavity (Fig.~\ref{SIFig:zipper}e) leading to a large optomechanical coupling for the fundamental in-plane mechanical mode of the bender-zipper structure.

\section{Optical background removal}

\begin{figure*}[ht]
	\includegraphics[scale=0.9]{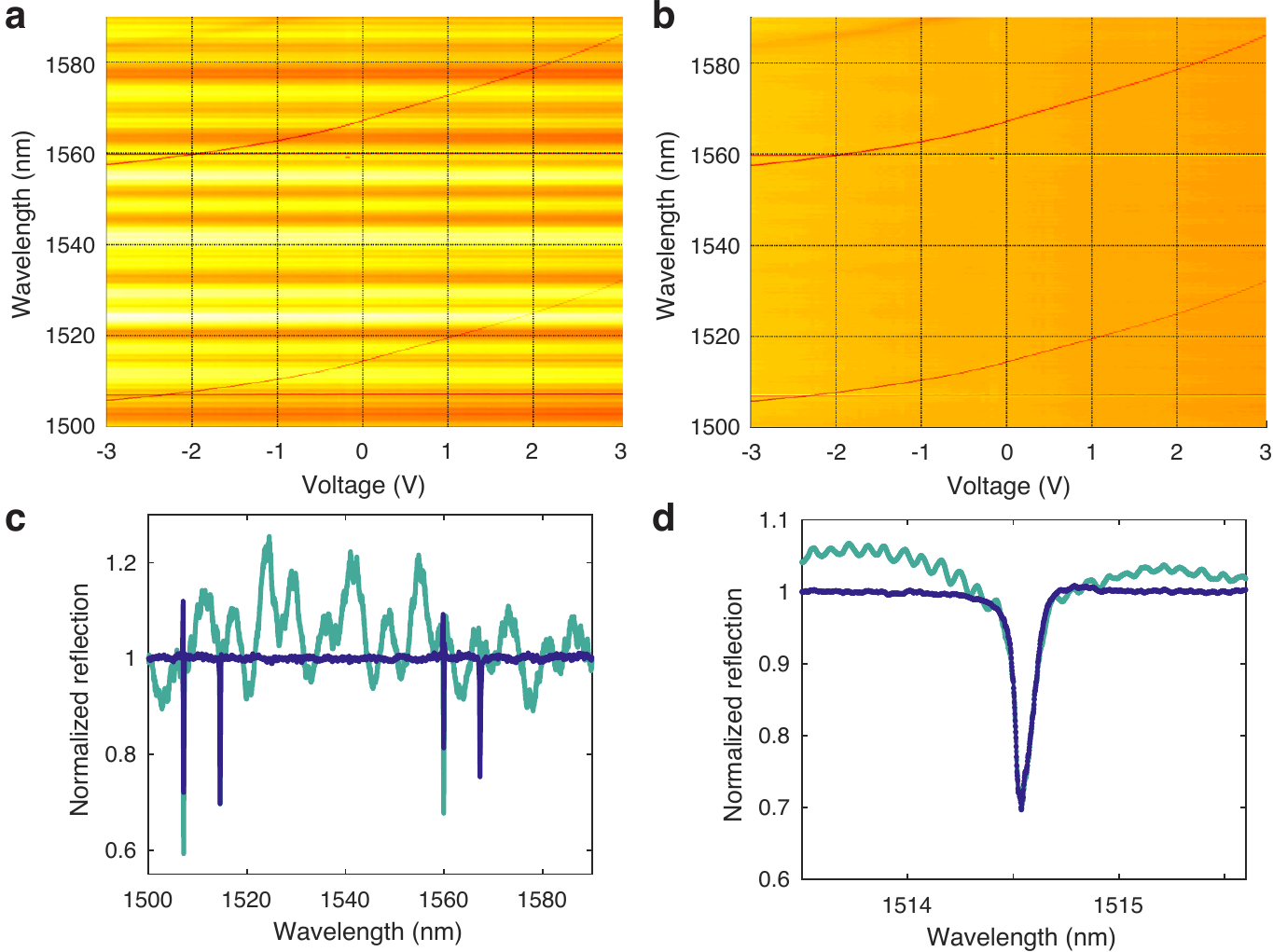} 
	\caption{\label{SIFig:background_removal}\textbf{Optical background removal in the presence of a tunable mode. a}, Raw measurement. \textbf{b}, Normalized measurement with background removal. \textbf{c}, Line-cut of (a) and (b) at $0$ V. \textbf{d}, Close-up of (c) around the fundamental optical mode of the active zipper cavity.}
\end{figure*}

In this section, we describe how we remove wavelength-dependent background fluctuations in optical reflection measurements that involve DC tuning. These fluctuations arise in the measurement setup and are not part of the actual device response. An example of raw DC tuning measurement data is shown in figure \ref{SIFig:background_removal}a. Even though it is still possible to see tuning of the mode, the background makes it difficult to process the data. Additionally, modes that are very weakly coupled can be hidden by the background. To remove the fluctuations in the background we use the fact that the resonance wavelength of the mode tunes with voltage whereas the background is static. For each wavelength, we can thus take the mean of the reflection over the DC voltages. Formally, for a reflection $r_{ij}$ at wavelength $i$ and voltage $j$, we compute the mean $m_i = \frac{1}{M}\sum_{j} r_{ij}$ where $M$ is the number of voltages. We then normalize the reflection $r_{ij,\text{norm}} = r_{ij}/m_i$ and recover the plot shown in Fig.~\ref{SIFig:background_removal}b. The fluctuating background is removed whereas the tunable modes are not affected. In Fig. \ref{SIFig:background_removal} we show a comparison of the raw and normalized reflection spectrum at $0$ V and see that the normalized background is flat. Because the control optical mode is static, we expect it to be removed through normalization. However, it is still visible, although slightly distorted. By looking at the control mode more carefully, we observe that it is slightly drifting with time as the experiment is being carried on, which means it is not perfectly static. Looking at a close up of one of the tunable modes (Fig.~\ref{SIFig:background_removal}d), we see that neither the wavelength nor the linewidth of the mode are affected by the background removal.

\section{Low temperature and pressure measurements}

\begin{figure*}[ht]
	\includegraphics[scale=0.95]{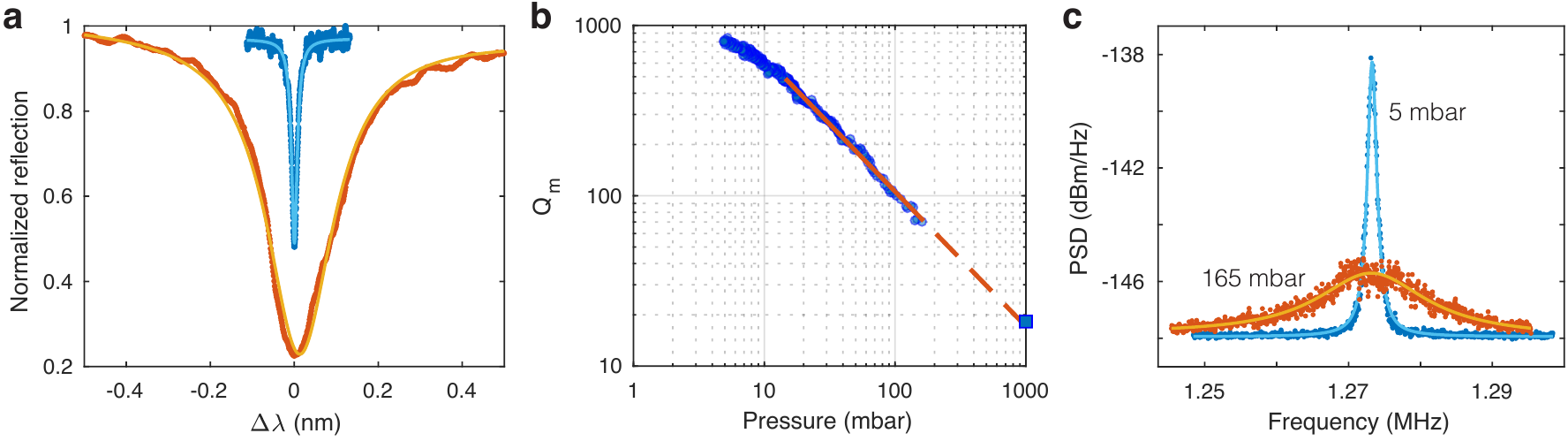} 
	\caption{\label{SIFig:lowT}\textbf{Low temperature and low pressure measurements. a}, Room temperature (red) and low temperature (blue) optical reflection spectrum of a bender-zipper cavity. The spectrum is fitted with a Lorentzian function to extract the linewidth in both cases. When the temperature is lowered, the mode is blue-shifted due to thermal contraction of the cavity. This shows that the optical mode is thermally broadened at room temperature. \textbf{b}, Q of a thermal mechanical resonance as a function of pressure at room temperature (blue circles). The red line is a fit, which when extrapolated (dashed red) to $1000$ mbar agrees well with a measurement performed outside of the vacuum chamber (blue square). \textbf{c}, Thermal power spectral density of one of the mechanical resonances at two different pressures. We use a Lorentzian function to fit the curves. Notice the narrowing of the mode at lower pressure due to reduced air damping.}
\end{figure*}

\subsection{Thermal broadening}
In this section, we show that the optical linewidths of the bender-zipper cavities are limited by thermal-mechanical broadening at room temperature. This effect causes the Lorentzian lineshape of the resonance to be convoluted by a Gaussian distribution representing the random position of the mechanical degree of freedom due to Brownian motion. The resulting profile is the so-called Voigt profile \cite{winger2011chip}. If the linewidth of the Gaussian $\kappa_\text{G}$ is much larger than the linewidth of the cavity resonance $\kappa$, then the linewidth of the Voigt profile is
\begin{equation}
\label{SIeqn:voigt}
\kappa_\text{V}\approx\kappa_\text{G}.
\end{equation}

To compute this broadened linewidth $\kappa_\text{G}$, we proceed with calculating the root mean square displacement $x_\text{rms}$ caused by thermal Brownian motion. We consider one half of a $L=$ \SI{7}{\micro\meter} bender-zipper cavity. By simulating the displacement profile of the fundamental mechanical in-plane mode, we calculate the effective motional mass associated with it through \cite{chan2012laser}: 
\begin{equation*}
	m_\text{eff} = \frac{\int dV\rho(\textbf{x})|\textbf{u}(\textbf{x})|^2}{\text{max}|\textbf{u}(\textbf{x})|^2} = 10.5 \text{ pg}.
\end{equation*}
Here, $\rho(\textbf{x})$ is the density of the material and \textbf{u}(\textbf{x}) is the displacement field. This mechanical mode has a frequency $\Omega_\text{m}/2\pi = 1.04$ MHz. Using the equipartition theorem, we find \cite{winger2011chip}: 
\begin{equation*}
	x_\text{rms,h} = \sqrt{\frac{k_B T}{m_\text{eff}\Omega_\text{m}^2}} = 95 \text{ pm}.
\end{equation*}
Here, $T = 293$ K is the room temperature and $k_B$ is the Boltzmann constant. The root mean square displacement of the full bender-zipper cavity is then simply given by: $x_\text{rms}=\sqrt{2}x_\text{rms,h} = 134$ pm. 
From optical resonance wavelength versus zipper cavity gap size  simulations (Fig.~\ref{SIFig:zipper}e), we extract the optomechanical coupling $G_\text{OM}/2\pi \approx 42$ GHz/nm. This value assumes a gap size of $\sim 100$ nm. Finally, the broadened linewidth is given by \cite{winger2011chip}:
\begin{equation*}
	f_G=2\sqrt{2\ln(2)}\cdot\frac{G_\text{OM}}{2\pi}\frac{\lambda^2}{c}x_\text{rms}=108~\text{pm}.
\end{equation*}
At a temperature of $4$ K, the same expression leads to $f_G = 13$ pm.
We therefore expect the linewidth of the cavity to decrease at low temperatures if it is limited by thermal broadening.

In Fig.~\ref{SIFig:lowT} we compare the measured reflection spectrum of a tunable bender-zipper cavity at room temperature and in a $\sim 4$ K cryostat. We observe that the linewidth decreases from $194$ pm to $16$ pm when cooling down which is more than an order of magnitude. The room temperature linewidth is on the same order as the theoretically predicted broadening. Because the gap $\tilde{g}_0$ is difficult to accurately measure, so is estimating $G_\text{OM}$. This could help explain the difference between predicted and measured values for the broadened linewidth. Additionally, simulating the full bender-zipper cavity as well as taking into account more mechanical modes that lead to broadening could help compute a more accurate value for the linewidth. In the low temperature case, $f_G$ is comparable to $\kappa$ and so equation \ref{SIeqn:voigt} does not hold anymore. Moreover, the resonance wavelength gets blue-shifted from $1534$ nm to $1521$ nm when cooling down which is expected from thermal contraction and refractive index change. The measured bender-zipper cavity has a gap $\tilde{g}_0 \approx 100$ nm which we extract from a scanning electron micrograph taken before release. The device was cooled down using a closed-cycle Montana Instruments cryostat.

Finally, we compute the zero-point fluctuation for the full bender-zipper cavity which is given by \cite{chan2012laser}:
\begin{equation*}
	x_\text{zpf}=\sqrt{2}\sqrt{\frac{\hbar}{2m_\text{eff}\Omega_\text{m}}}\approx 39 \text{ fm}.
\end{equation*}
This allows us to obtain the vacuum optomechanical coupling strength $g_0 = G_\text{OM}\cdot x_\text{zpf} \approx 2\pi\times 1.6$ MHz.
\subsection{Mechanical quality factor as a function of pressure}
We investigate how the thermal-mechanical power spectrum of the bender-zipper cavity changes when pressure is decreased. At atmospheric pressures, the mechanical quality factor $Q_\text{m} \approx 18$ is limited by air damping. We use the cryostat from the previous section as a vacuum chamber at room temperature and measure the thermal-mechanical spectrum as the chamber is being pumped down. We extract $Q_\text{m}$ as a function of pressure (Fig.~\ref{SIFig:lowT}b and \ref{SIFig:lowT}c). As the pressure decreases, the quality factor increases at a slightly sub-linear rate. We observe a two-order of magnitude increase in the mechanical quality factor, reaching $Q_\text{m} \sim 1000$ at $5 $ mbar. Extrapolating this rate to atmospheric pressure at $1000$ mbar agrees well with measurement. Below $\sim 10$ mbar, we can see $Q_\text{m}$ start saturating, showing that air damping is no longer the main mechanism for dissipation. Thermo-elastic damping starts becoming dominant at room temperature and low pressure, so we expect $Q_\text{m}$ to further improve at low temperature.

\section{Rotated nanobenders}
We have considered $Y$-cut LN where the nanobender is parallel to the crystal $X$ axis. In this section we investigate how changing the in-plane orientation of the nanobender affects the displacement $\bm u$ of the nanobender. This amounts to changing the angle $\varphi$ between $X$ and $x$ (parallel to the nanobender), where $X$ is the crystal axis, $x$ is the global coordinate fixed with the nanobender and is pointing from the fixed-end towards the free-end of the nanobender. Global $z$ axis is perpendicular to the chip (parallel to crystal $Y$ axis).

\begin{figure*}[ht]
	\includegraphics[scale=0.95]{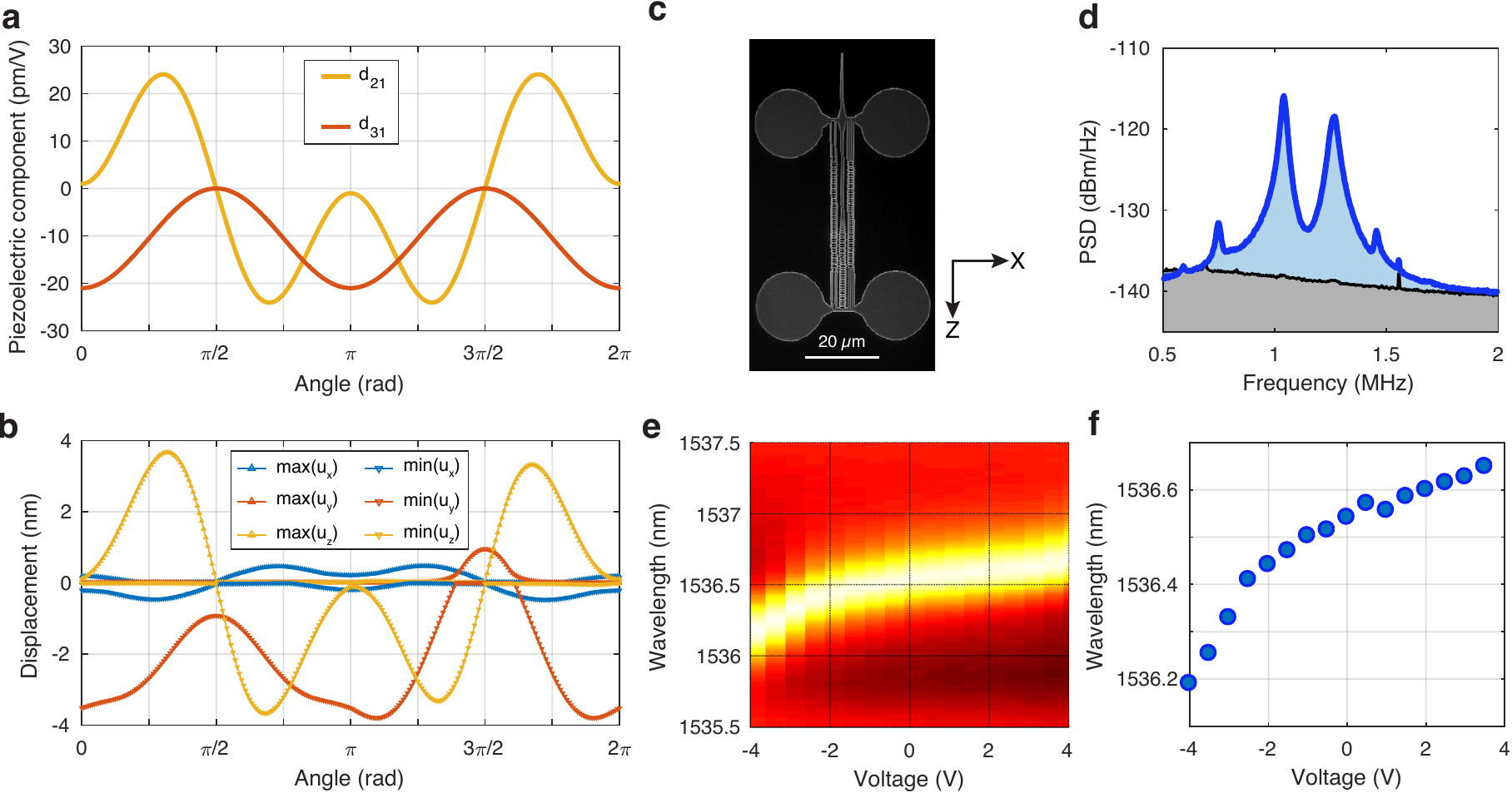} 
	\caption{\label{SIFig:rotated}\textbf{Bender-zipper cavity aligned along crystal axis $Z$. a}, Piezoelectric components for $Y$-cut lithium niobate versus angle between global axis $x$ and crystal axis $X$. \textbf{b}, Simulated displacement of a nanobender versus nanobender orientation. An angle of $0^\circ$ ($90^\circ$) corresponds to the nanobender parallel to crystal axis $X$ ($Z$). \textbf{c}, Scanning electron micrograph of a bender-zipper cavity aligned along crystal axis $Z$ (taken before the aluminum evaporation).  \textbf{d}, Thermal power spectral density of the rotated bender-zipper cavity. \textbf{e}, Measured DC tuning of the fundamental optical mode of a rotated bender-zipper cavity attached to $L=\SI{10}{\micro\meter}$ nanobenders through narrow tethers. The background is removed through normalization. \textbf{f}, Wavelength tuning as a function of voltage. The data is extracted from Fig.~\ref{SIFig:rotated}e.}
\end{figure*}

A good indicator for how the displacement varies with $\varphi$ is to look at how the rotated piezoelectric components $d_{21}$ and $d_{31}$ change with $\varphi$ (Fig.~\ref{SIFig:rotated}a). As discussed in Sec.~\ref{SISubsec:derive-single-bender}, $d_{21}$ ($d_{31} $) couples between $E_y $ ($E_z $) and $S_{xx}$, and generates vertical (horizontal) bending from $\partial_z E_y $ ($\partial_y E_z $). We can directly compare it to simulated displacements of a nanobender with $L=$ \SI{10}{\micro\meter} where $\varphi$ is swept from $0$ to $2\pi$. 

Plotting the maximal and minimal value of the displacement (Fig.~\ref{SIFig:rotated}b), we see that the out-of plane component $u_z$ varies approximately as $d_{21}$. The in-plane displacement component perpendicular to the nanobender $u_y$ follows $d_{31}$ relatively well when $\varphi$ is changed, although the resulting curves are asymmetric. We attribute this to one end of the nanobender being anchored, breaking the symmetry, and also contributions from other non-zero piezoelectric components. When $\varphi = 90^\circ$, we see that the in-plane displacement $u_y$ is reduced by a factor $4.5$. Moreover, for a bender-zipper cavity, two nanobenders rotated by $\pm 90^\circ$ are attached to the two ends of the same half of the zipper cavity. The two nanobenders would bend towards opposite directions (Fig.~\ref{SIFig:rotated}b, at $\pi/2 $ and $3\pi/2$) if they were not connected, leading to a small net displacement. From simulations of a rotated bender-zipper cavity, we find that $u_y$ decreases by more than one order of magnitude.

We fabricate and measure such a bender-zipper cavity where $\varphi = 90^\circ$. The device (Fig.~\ref{SIFig:rotated}c) does not have a waveguide with a $90^\circ$ bend as opposed to the bender-zipper cavity aligned along crystal axis $X$ in the main text. Moreover the nanobenders are attached to the zipper cavity through small tethers. We confirm that the active bender-zipper cavity is free to move by measuring its thermal-mechanical power spectral density (Fig.~\ref{SIFig:rotated}d) which looks typical of previously measured tunable devices. As expected, the DC tuning measurement of the rotated device (Fig.~\ref{SIFig:rotated}e and f) shows only little tuning. We extract a DC tuning coefficient of $\alpha \sim 0.04$ nm/V which is significantly smaller than devices with $\varphi = 0$ where we measure $\alpha \sim 1$ nm/V. Moreover, we measure that the control zipper cavity does not tune at all.

Lastly it is worth noting that according to the simulation results, a variety of actuation directions can be achieved with the same crystal cut by fabricating the nanobender along different in-plane orientations. For example, relatively clean vertical actuation can be achieved at $\varphi \approx 270 \pm 20$ degrees even on $Y$-cut LN.

\section{Mechanical mode frequency and AC modulation measurement}

\subsection{Fundamental mechanical mode frequency as a function of nanobender length}
\label{SIsubsec:mech-freq}
\begin{figure*}[ht]
	\includegraphics[scale=0.95]{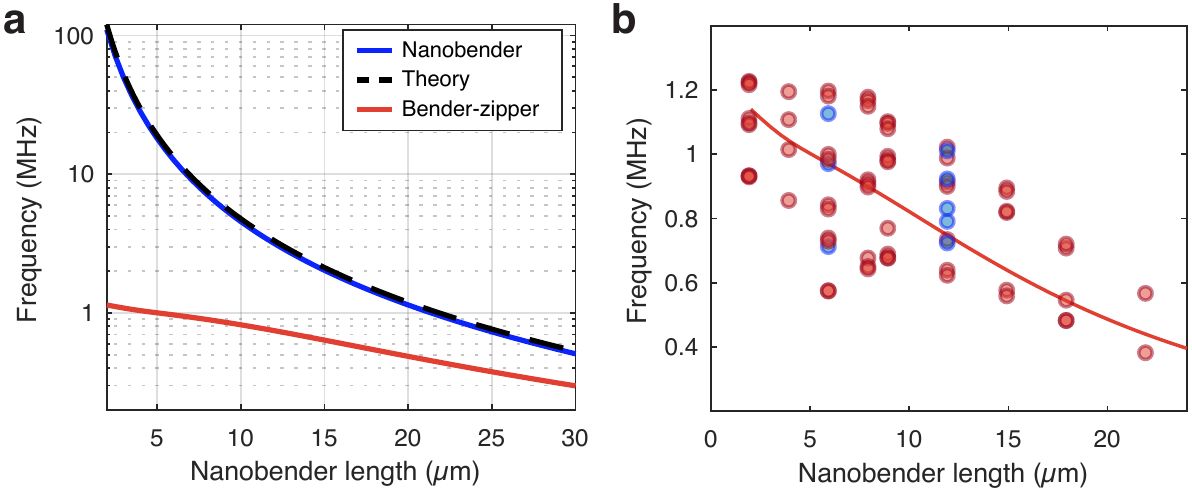} 
	\caption{\label{SIFig:mech_freqs}\textbf{Mechanical resonance frequencies. a}, Finite element method simulation of the eigenfrequencies of a nanobender only (blue) and a bender-zipper cavity (red). The dashed black curve is calculated from Euler-Bernoulli beam theory. \textbf{b}, Measurement of the resonance frequencies of $21$ bender-zipper cavities with varying nanobender lengths $L$. Devices without and with tether (blue and red circles) are plotted. The red line is the same as the one in Fig.~\ref{SIFig:mech_freqs}a.}
\end{figure*}

In this section, we study the actuation speed of a single nanobender as well as the tuning speed of the bender-zipper cavity. For the case of a single nanobender, Euler-Bernoulli beam theory allows us to write a simple expression for its fundamental eigenfrequency. The in-plane fundamental eigenfrequency of a beam clamped on one side, is given by \cite{cleland2013foundations} 
\begin{eqnarray*}
	f_1 &\approx& \frac{3.5161}{2\pi L^2}\sqrt{\frac{EI}{\mu}}\\
	&=& \frac{3.5161}{2\pi}\sqrt{\frac{E}{12\rho}}\frac{w}{L^2}
\end{eqnarray*}
where $E$ is Young's modulus, $I=\frac{tw^3}{12}$ is the second moment of area and $\mu = \rho w t$ is the mass per unit length. The density of LN is $\rho = \SI{4700}{\kilo\gram\per\cubic\meter}$. The agreement with finite element method simulations (see Fig.~\ref{SIFig:mech_freqs}a) is very good, overestimating the actual values by only $\sim 5\%$. This deviation arises because the analytical expression is not taking into account the anisotropy of LN (we approximated LN as an isotropic material and took $E \approx 2.03 \cdot 10^{11}$ Pa \cite{weis1985lithium}). Furthermore, the analytical expression does not include the aluminum electrodes which would decrease the frequency due to the much smaller Young's modulus of aluminum. We can recover the spring constant of the nanobender: $k=4\pi^2\mu L f_1^2 \approx 5.7$~N/m for $L=\SI{10}{\micro\meter}$. This lets us convert the displacement actuation to an equivalent force per volt of $F/U =k\Delta /U\approx 5.7\text{ N/m}\cdot 3.5 \text{ nm/V}\approx 20\text{ nN/V}$. Because $k$ scales as $L^{-3}$ and $\Delta /U$ scales as $L^2$, the force per volt scales as $L^{-1}$. 

In Fig.~\ref{SIFig:mech_freqs}a, we also show the simulated fundamental resonance frequency of one half of a bender-zipper cavity as a function of the nanobender length $L$. As expected, for small $L$, the nanobender is not limiting the tuning speed of the cavity but rather the size of the cavity itself. In Fig.~\ref{SIFig:mech_freqs}b, we compare this simulated curve to experimental values obtained from thermal mechanical spectra of $21$ bender-zipper cavities. For each device, we observe multiple resonance frequencies which are spread around the curve obtained from simulation. Simulating the full structure should lead to a splitting of the resonance frequency due to the mechanical coupling between the two identical halves of the bender-zipper cavity.

\subsection{AC wavelength shift as a function of modulation voltage}
\begin{figure*}[ht]
	\includegraphics[scale=0.95]{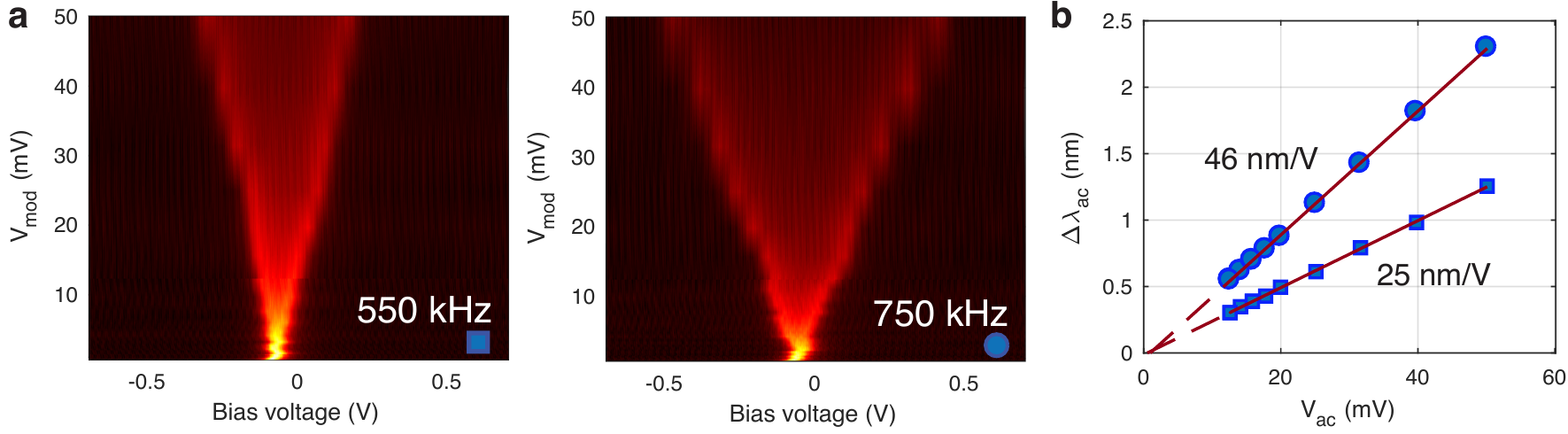} 
	\caption{\label{SIFig:alpha_vpp}\textbf{Sweeping the AC voltage applied to a bender-zipper cavity. a}, Modulation measurement results for two different modulation frequencies corresponding to two different mechanical resonances. The device measured here is the same as the one in Fig.~3c in the main text. \textbf{b}, Analytical fit (blue circles) of both modulation measurements shown in Fig.~\ref{SIFig:alpha_vpp}a. As explained in the main text, bias voltage is converted to wavelength shift. The red line is a linear fit.}
\end{figure*}

Here, we experimentally confirm that the AC wavelength shift $\Delta\lambda_\text{ac}$ of the bender-zipper cavity is linear with respect to the driving voltage $V_\text{ac}$. In Fig.~\ref{SIFig:alpha_vpp}a, we show results of modulation experiments where we sweep $V_\text{ac}$ from $0$ to $50$ mV. This measurement is done for two different resonance frequencies of the bender-zipper cavity (see Fig.~3e in the main text). We proceed as in the main text and extract $\Delta\lambda_\text{ac}$ as a function of $V_\text{ac}$ by doing an analytical fit and converting bias voltage to wavelength. These are plotted in Fig.~\ref{SIFig:alpha_vpp}b. By fitting a line, we extract the AC tuning coefficient for a specific resonance frequency. We find that the AC tuning coefficient is independent of $V_\text{ac}$ for the range of voltages considered here, validating the voltage normalization done in Fig.~3e of the main text.

\end{document}